\documentclass[leqno,12pt,a4paper]{article}
\usepackage{latexsym}
\usepackage{amsmath}
\usepackage{amssymb}
\usepackage{bbm}
\usepackage{amsthm}
\usepackage[misc]{ifsym}

\usepackage[]{algorithm,algpseudocode}
\algnewcommand{\IIf}[1]{\State\algorithmicif\ #1\ \algorithmicthen}
\algnewcommand{\IElse}{\unskip\ \algorithmicelse\ }
\algnewcommand{\EndIIf}{\unskip\ \algorithmicend\ \algorithmicif}

\usepackage{graphicx}
\usepackage{epsfig}

\usepackage{tikz}
\usetikzlibrary{shapes,arrows}

\allowdisplaybreaks

\usepackage{tikz}
 
\newcommand{\boldface}[1]{\boldsymbol{#1}}
\newcommand{\bfa}{\boldface{a}}
\newcommand{\bfb}{\boldface{b}}

\newcommand{\bff}{\boldface{f}}

\newcommand{\bfl}{\boldface{l}}

\newcommand{\bfn}{\boldface{n}}

\newcommand{\bfp}{\boldface{p}}
\newcommand{\bfq}{\boldface{q}}

\newcommand{\bft}{\boldface{t}}
\newcommand{\bfu}{\boldface{u}}
\newcommand{\bfv}{\boldface{v}}

\newcommand{\bfx}{\boldface{x}}

\newcommand{\bfC}{\boldface{C}}
\newcommand{\bfD}{\boldface{D}}

\newcommand{\bfF}{\boldface{F}}

\newcommand{\bfI}{\boldface{I}}

\newcommand{\bfL}{\boldface{L}}
\newcommand{\bfM}{\boldface{M}}

\newcommand{\bfP}{\boldface{P}}

\newcommand{\bfS}{\boldface{S}}

\newcommand{\bfX}{\boldface{X}}

\newcommand{\bfalpha}{\boldsymbol{\alpha}}

\newcommand{\bfvarepsilon}{\boldsymbol{\varepsilon}}

\newcommand{\bfsigma}{\boldsymbol{\sigma}}

\newcommand{\bfupsilon}{\boldsymbol{\alpha}} 
\newcommand{\bfeps}{\boldsymbol{\varepsilon}}
\newcommand{\calA}{\mathcal{A}}

\newcommand{\calD}{\Delta}
\newcommand{\calE}{\mathcal{E}}

\newcommand{\calG}{\mathcal{G}}
\newcommand{\calH}{\mathcal{H}}

\newcommand{\calK}{\mathcal{K}}
\newcommand{\calL}{\mathcal{L}}

\newcommand{\calP}{\mathcal{P}}
\newcommand{\calQ}{\mathcal{Q}}

\newcommand{\calW}{\mathcal{W}}

\usepackage{bbold}

\newcommand{\dsC}{\mathbb{C}}

\newcommand{\be}{\begin{equation}}
\newcommand{\ee}{\end{equation}}
\newcommand{\bea}{\begin{eqnarray}}
\newcommand{\eea}{\end{eqnarray}}
\newcommand{\bes}{\begin{equation*}}
\newcommand{\ees}{\end{equation*}}
\newcommand{\beas}{\begin{eqnarray*}}
\newcommand{\eeas}{\end{eqnarray*}}
\newcommand{\D}{\displaystyle}
\newcommand{\dd}{\ \mathrm{d}}

\newcommand{\intO}{\int_\Omega}
\newcommand{\intPO}{\int_{\partial\Omega}}
\newcommand{\into}{\int_\omega}
\newcommand{\intPo}{\int_{\partial\omega}}
\newcommand{\nablax}{\nabla_{\bfx}}
\newcommand{\nablaX}{\nabla_{\bfX}}
\newcommand{\Dt}[1]{\frac{\mathrm{D}#1}{\mathrm{D}t}}

\newcommand{\pf}[2]{\frac{\partial #1}{\partial #2}}
\newcommand{\comp}[1]{{(#1)}}
\newcommand{\refc}{\comp{0}}

\definecolor{RUBblue}{rgb}{0.0470588,0.262745,0.411765}
\definecolor{lightgray}{gray}{0.92}
\definecolor{blau}{rgb}{0,0.25,1.0}

\def\dDs{{
D^\text{s}}}
\def\ddDs{{
\Delta^\text{s}}}

\def\ddDsrefc{{
\Delta^{\text{s}}}}
\def\ddDsrefcRD{{
\Delta^{\text{s}}_\text{RD}}}
\def\ddDsrefcRI{{
\Delta^{\text{s}}_\text{RI}}}
\newcommand{\diss}{\ddDs} 
\newcommand{\dissrefc}{\ddDsrefc} 
\newcommand{\dissI}{\dDs} 

\newcommand{\bfpdissrefc}{\bfp^\text{s}}
\newcommand{\bfpdissT}{\bfp^\text{diss}}
\newcommand{\bfpdissTrefc}{\bfp^\text{diss}}

\newcommand{\graybox}[1]{\colorbox{gray!30}{$#1$}}
\newcommand{\Graybox}[1]{\begin{tcolorbox}[breakable, enhanced jigsaw] #1 \end{tcolorbox} }%
\newcommand{\Pidissrefc}{\calD^{\text{diss}}}

\usepackage[most]{tcolorbox}

\newcommand{\el}{\text{e}}
\newcommand{\pl}{\text{p}}
\newcommand{\bfFp}{\tilde{\bfF}{}^\pl }
\newcommand{\dbfFp}{\dot{\tilde{\bfF}}{}^\pl }

\newcommand{\bfepsV}{\bfeps^\text{v}}
\newcommand{\dbfepsV}{\dot{\bfeps}^\text{v}}

\newcommand{\cint}{c_\text{int}}

\newcommand{\powerM}{\re{\calP^\text{mech}}}
\newcommand{\powerT}{\re{\calP^\text{therm}}}
\newcommand{\heatflux}{\bfq}
\newcommand{\heatsource}{h}
\newcommand{\entropyflux}{\bfq^\text{s}}
\newcommand{\entropysource}{\bar{h}^\text{s}}

\usepackage{lipsum}
\usepackage{placeins}

\definecolor{RUBblue}{rgb}{0.0470588,0.262745,0.411765}

\newcommand{\re}[1]{#1}

\newcommand{\rev}[2]{#2}

\usepackage{booktabs}

\begin{document}

\title{An extended Hamilton principle as unifying theory for coupled problems and dissipative microstructure evolution}
\date{} 
\maketitle
{\large
\noindent Philipp Junker\re{$^1$}, Daniel Balzani\re{$^2$}\\[0.5mm]
}
\re{$^1$: Leibniz University Hannover, Institute of Continuum Mechanics,  Hannover, Germany}\\
\re{$^2$: }Chair of Continuum Mechanics, Ruhr University Bochum, Bochum, Germany\\[2mm]
{
Corresponding author:\\[0.5mm]
Daniel Balzani, \color{RUBblue}{\Letter \hskip 1mm daniel.balzani@rub.de} 
}


\section*{Abstract}
An established strategy for material modeling is provided by energy-based principles such that evolution equations in terms of ordinary differential equations can be derived. However, there exist a variety of material models that also need to take into account non-local effects to capture microstructure evolution. In this case, the evolution of microstructure is described by a partial differential equation. In this contribution, we present how Hamilton's principle provides a physically sound strategy for the derivation of transient field equations for all  state variables. Therefore, we begin with a demonstration how Hamilton's principle generalizes the principle of stationary action for rigid bodies. Furthermore, we show that the basic idea behind Hamilton's principle is not restricted to isothermal mechanical processes. In contrast, we propose an extended Hamilton principle which is applicable to coupled problems and dissipative microstructure evolution. As example, we demonstrate how the field equations for all state variables for thermo-mechanically coupled problems, i.e. displacements, temperature, and internal variables, result from the stationarity of the extended Hamilton functional. The relation to other principles, as principle of virtual work and Onsager's principle, are given. Finally, exemplary material models demonstrate how to use the extended Hamilton principle for thermo-mechanically coupled \rev{28}{elastic,} \rev{1}{gradient-enhanced, rate-dependent, and rate-independent}  materials. 

\section{Introduction}
Physical bodies are described by a varying number of state variables. For instance, two state variables are necessary in a thermo-mechanical sense: displacements $\bfu$ and temperature $\theta$. Other problems could be chemo-mechanical ones with the displacements and the concentration of substances, electro-mechanical ones with the displacements and the electric field, or combinations of the latter. Particularly for solids,  these state variables, however, do not carry enough information to reflect the total state. The important influence of microstructure and its time-dependent configuration is not accounted for, i.e. for thermo-mechanically coupled problems, by the primary state variables $\bfu$ and $\theta$. To overcome this deficiency, additional variables for the ``inner state'' were introduced between 1820 and 1850 by Carnot, Joule, and Clausius \cite{carnot2012reflections,joule1843philosophical,clausius1850bewegende}. More details on the history is given in \cite{horstemeyer2010historical}. These variables are nowadays referred to as internal variables which might be denoted, arbitrarily, by the vector $\bfupsilon$. Then, the set $\{\bfu,\theta,\bfupsilon\}$ completely describes the physical state of thermo-mechanically coupled problems and dissipative mircostructure evolution. Appropriate choices for $\bfupsilon$ mimic the experimentally observed material configuration. However, a sophisticated challenge of modern mechanics is the development of mathematical equations that describe how the microstructural state, given in terms of $\bfupsilon$, evolves. Consequently, these equations are referred to as evolution equations and form the material model. Herein, a material-specific choice of $\bfupsilon$ defines answers to questions as is it a plastic strain, a volume fraction, a hardening or damage variable, or others.

Along with the choice which set of internal variables shall be used in the material model, the identification of constraints from experimentally observed material behavior is a crucial step during the derivation of a related mathematical description of the material behavior. Then, the consideration of the constraints in terms of mathematical formulas often turns out to be a non-trivial issue. It is thus convenient to derive material models from a potential which might comprise several benefits: ensurance of thermodynamic consistency and simplified consideration of constraints by using appropriate Lagrange parameters. 
One potential-based principle of material modeling is the principle of maximum dissipation. Here, the dissipation, being a function of the thermodynamic driving forces $\bfp$, is maximized to derive the evolution equations. The assumption that (microstructural) processes maximize the dissipation is physically sound and, consequently, material models that are derived by using this approach agree to the fundamental laws of physics.
A similar potential-based principle is given by the so-called principle of the minimum of the dissipation potential. In contrast to the principle of maximum dissipation, the potential is here a function of the thermodynamic flux $\dot{\bfupsilon}$ such that minimizing the dissipation potential directly results in the evolution equations. A subsequent Legendre transformation of the associated dissipation function \rev{25}{allows to identify the related yield} function $\Phi$ if rate-independent processes are modeled.

It is obvious that a model for a specific material should be able to describe the related processes with sufficient accuracy and, of course, must not contradict fundamental physical observations and their related laws. These requirements still provide a huge variety to construct a specific material model. For instance, the relation
\be
\dot{\bfupsilon} \propto \bfp
\ee
is often used for rate-dependent processes. This proportionality between the thermodynamic driving force $\bfp$, which can be derived from the free energy density $\Psi$ by
\be
\bfp := - \pf{\Psi}{\bfupsilon} \ ,
\ee
and the thermodynamic flux $\dot{\bfupsilon}$ routes back to Onsager and is consequently known as Onsager's principle. In case of rate-independent processes, materials are modeled by defining a suitable \rev{25}{yield} function $\Phi$, and the evolution of the internal variable is assumed to be obtained from a thermodynamic extremal principle, e.g. principle of maximum dissipation or principle of the minimum of the dissipation potential, yielding
\be
\dot{\bfupsilon} \propto \pf{\Phi}{\bfp}
\ee
This procedure has proven to be beneficial when quasi-static processes evolve. Whereas for appropriate choices for $\Phi$ and using Onsager's principle it can be shown that such material models are thermodynamically sound by construction, the verification of the latter is usually a part of the derivation of a material model if other modeling techniques are employed. Here, the Coleman-Noll procedure \rev{3}{is used for the identification of the driving forces and thermodynamic consistency is proved by checking the dissipation inequality
\be
\bfp\cdot\dot{\bfalpha} \ge 0 \ .
\ee}%
However, a drawback of these potential-based approaches is that they are not suitable for the derivation of field equations for the internal variable. To be more precise, the inclusion of gradient terms is not possible since the principle of maximum dissipation and the principle of the minimum of the dissipation potential are both defined locally for a material point. A solution to this drawback is given by Hamilton's principle: it is formulated as variational problem for the entire physical body. Thus, field equations, i.e.~partial differential equations in time and space, can be derived from related stationarity conditions. Hamilton's principle is strongly related to the principle of stationary action and comprises the principle of virtual work and the principle of the minimum of the dissipation potential, which all turn out to be special cases of Hamilton's principle.

In this contribution, we aim at highlighting the relation of Hamilton's principle to the fundamental laws of thermodynamics and other extremal principles of continuum mechanics. We begin with a short overview on the history of material modeling and give a brief introduction to thermodynamics. Later, we recall the principle of stationary action and demonstrate its relation to the 1$^\text{st}$ law of thermodynamics. This serves as basis to present how a generalization of the principle of stationary action, which is originally restricted to rigid particles, can be used to describe non-conservative deformable solids. Particularly, we show that an extended Hamilton principle thus might be used for the derivation of material models including field equations for all state variables in thermo-mechanically coupled problems given by $\{\bfu,\theta,\bfupsilon\}$. Finally, we show basic examples of thermo-mechanically coupled material models based on the extended Hamilton principle.

\section{Brief outline of the history of material modeling}
Material modeling comprises the usage of internal variables and strategies to derive governing equations for describing their evolution. A prominent modeling strategy is the application of variational methods which were invented far earlier than the concept of internal variables, see \cite{horstemeyer2010historical}. The idea of some quantity that is maximized or minimized by physical processes roots back to Aristotle or even before, cf. \cite{berdichevsky2009variational}. It is strongly related to the principle of stationary action which was embedded into a rigorous mathematical concept in the late 1700's by Leibniz, Euler, Maupertuis, Lagrange, and others, see \cite{pulte1989prinzip}. The invention of variational calculus allowed to solve the problem of brachistochrone as well as light refraction, the motion of a conservative oscillator, and many more. A related approach is the well-known principle of Castigliano and Menabrea \cite{menabrea1868etude,castigliano1875nuova}. A historical review is presented in \cite{capecchi2010historical}. Hamilton generalized the principle of stationary action in \cite{hamilton1834general,hamilton1835second}. More details on Hamilton's principle are given by Bedford \cite{bedford1985hamilton} and Bailey \cite{bailey2002unifying} who presents a clear distinction of different energy-based extremal principles.

The usage of internal variables in current sense routes back to the concept of ``variables of inner state'' as pioneered by Carnot \cite{carnot2012reflections}, Joule \cite{joule1843philosophical}, and Clausius \cite{clausius1850bewegende}. Afterwards, the concept was put in a sound thermodynamic framework by Helmholtz \cite{von1887ueber}, Thomson \cite{thomson1853xv}, Maxwell \cite{maxwell1875dynamical}, Gibbs \cite{gibbs1873method}, and Duhem \cite{duhem1991aim} (see \cite{horstemeyer2010historical} and \cite{maugin2015saga} for a detailed historical presentation). In 1931, Onsager used the thermodynamic approach of thermodynamic fluxes $\dot{\bfupsilon}$ and thermodynamic forces $\bfp$. Starting from the (reduced) Clausius-Duhem inequality 
\be
\bfp \cdot \dot\bfupsilon \ge 0 \ ,
\ee
he postulated a proportionality which served as fundamental \rev{4}{ansatz for material modeling since it fulfills thermodynamic consistency a priori} \cite{onsager1931reciprocalI,onsager1931reciprocalII}. Roughly at the same time, the concept of thermodynamic fluxes and forces was conveyed to continuum mechanics by Bridgman \cite{bridgman1936nature} and Eckart \cite{eckart1940thermodynamics}.  A time-dependent evolution was investigated in the general work by Ziegler \cite{ziegler1958attempt}. The thermodynamics of irreversible processes was put in the 1960's into a more rigourous mathematical framework by Truesdell \cite{truesdell2012rational} and Coleman and Noll \cite{coleman1963thermodynamics,coleman1964thermodynamics,
coleman1967thermodynamics} who established so-called rational thermodynamics. A further elaboration by Rice \cite{rice1971inelastic} and Lubliner \cite{lubliner1973structure} founded in the 1970's the concept of internal variables in its current form. The concept of maximum dissipation as fundamental strategy for the derivation of evolution equations was given by Lubliner in 1984 \cite{lubliner1984maximum} and Simo in 1988 \cite{simo1988framework}.~A concept for complex microstructures was proposed by  Ortiz and Repetto in their work on non-convex energy minimization \cite{ortiz1999nonconvex} which built upon the mathematical basis of Ball and James \cite{ball1989fine}. An alternative for the principle of maximum dissipation was introduced by the principle of the minimum of the dissipation potential, published in various forms e.g.~by Martin, Maier, Halphen, Maugin and Hackl \cite{martin1965note,maier1969some,halphen1975materiaux,
maugin1992thermomechanics,hackl1997generalized}. A comparison between the two approaches was given in \cite{hackl2008relation}. \rev{a}{Strategies to couple the local material behavior to non-local fields can be given by the introduction of penalty terms, see e.g.~\cite{dimitrijevic2011regularization}, or by postulating evolution equations that are added as constraints to the functional by using Lagrange parameters, see e.g. the works by Hackl and Fischer~\cite{hackl2011study,fischer2016improved}.  } 
\rev{35 / a}{It is also worth mentioning that instead of a time-continuous presentation time-incremental variational methods for material modeling have been established. Examples are given in a more numerical setting by Simo~\cite{simo1990variational}. A more analytical setting is discussed by Mielke for the case of finite elasto-plasticity in~\cite{mielke2003energetic} based on dissipation distances and by Miehe who generalized this approach to gradient-extended standard dissipative materials~\cite{miehe2011multi}. More details are also provided in the textbook by Mielke in 2015~\cite{mielke2015rate}.}

\section{Thermodynamic basis}
In this section we recapitulate shortly the fundamentals of thermodynamics to define our notation and set the basis for the description of the generalization of Hamilton's principle. Here, we restrict ourselves to thermo-mechanically coupled processes. 

\subsection{The first law of thermodynamics}
The first law of thermodynamics postulates the balance of energy which states that the rate of the total energy of the body equals the power due to mechanical and thermal loads. This can be recast in the equation
\be
\label{eq:BalEnergyGlob}
\dot{\calE} + \dot{\calK} = \powerM + \powerT
\ee
where the internal energy is given by $\calE$ and the kinetic energy by $\calK$; the mechanical power is given by $\powerM$ and the thermal power by $\powerT$, cf. \cite{altenbach2012kontinuumsmechanik}. The same axiom is given, equivalently, after integrating \eqref{eq:BalEnergyGlob} over time by
\be
\calE+\calK = \calW + \calQ + c_\text{int}
\ee
with \rev{10}{the mechanical work $\calW$, the thermal work $\calQ$, and} some integration constant $c_\text{int}$. The internal energy \rev{6}{of the full body} $\calE$ consists of the \rev{6}{total} free energy $\into\Psi \dd v$ and the heat $\into h^\text{in} \dd v$, i.e., energy contributions which are neither related to external work nor mass transport. The heat is thus the amount of energy stored in the thermal movement of the atoms and measured by $h^\text{in} = \rho \theta \bar{s} $ with the absolute temperature $\theta$ and the mass-specific entropy $\bar{s}=s/\rho$. The entropy can accordingly be interpreted as temperature-specific energy which, according to the 2$^\text{nd}$ law of thermodynamics, indicates the part of the internal energy that is not directly accessible by mechanical processes. The internal energy is then given by
\be
\label{eq:DefE}
\calE = \into \rho\, \bar{\Psi} \dd v + \into \rho\,\theta \bar{s} \dd v = \int_\Omega \rev{5}{\rho} \, \bar{\Psi}  J \dd V + \int_\Omega \rev{5}{\rho} \, \theta \bar{s} J \dd V \ .
\ee
Here, $\bar{\Psi}$ denotes the mass-specific free energy density and $\rho$ the mass density. The material domain in the current configuration is termed as $\omega$ whereas the material domain in the reference configuration is indicated by $\Omega$. The transformation of the volume element in the current configuration to its equivalence in the reference configuration is performed by $\dd v = J \dd V$
with $J:= \mathrm{det}\bfF$ and the deformation gradient $\bfF:=\bfI + \partial \bfu/\partial \bfX$ with the identity tensor of second order $\bfI$, see \cite{altenbach2012kontinuumsmechanik}. The spatial coordinate in the current configuration is denoted as $\bfx$ and in the reference configuration as $\bfx(t=0)=\bfx^\comp{0}=:\bfX$. The free energy can be expressed by
\be
\label{eq:DefPsi}
\into \Psi \dd v = \calE - \into \theta s \dd v 
\ee
while the kinetic energy is given by
\be
\label{eq:DefKin}
\calK = \into \frac{1}{2} \rho\, ||\dot{\bfu}||^2 \dd v
\ee
with the velocity $\dot{\bfu}$. External forces can be divided into surface and body forces, indicated by the traction vector $\bft^\star$ at the boundary of the body $\partial\omega$ and mass-specific body force $\bar{\bfb}^\star$. Then, the mechanical work $\calW$ is given by
\be
\label{eq:DefW}
\calW = \intPo \bft^\star\cdot\bfu \dd a + \into \rho\,\bar{\bfb}^\star\cdot\bfu \dd v 
\ee
whereas the mechanical power reads
\be
\label{eq:DefWd}
\powerM = \intPo \bft^\star \cdot \dot{\bfu} \dd a + \into  \rho\,\bar{\bfb}^\star \cdot\dot{\bfu} \dd v = \into \bfsigma:\bfD \dd v + \into \rho \, \frac{\mathrm{D}}{\mathrm{D}t} ||\dot{\bfu}||^2 \dd v \ ,
\ee
see \cite{altenbach2012kontinuumsmechanik}. Here, $\bfsigma$ denotes the Cauchy stress; the stretch rate tensor $\bfD:=(\bfL+\bfL^T)/2$ is the symmetric part of the spatial velocity gradient $\bfL:=\partial \bfv/\partial \bfx$. \rev{17}{It is worth mentioning that we use pure Neumann boundary conditions, i.e., $\partial\omega=\Gamma_\sigma$ and $\Gamma_u=\emptyset$ where $\Gamma_\sigma$ denotes the boundary with prescribed tractions and $\Gamma_u$ denotes the boundary with prescribed displacements. However, the complete derivation which follows holds true when also Dirichlet boundary conditions are used. We skip the straightforward inclusion of Dirichlet boundary conditions here for a more convenient presentation.} The material time derivative is defined as
\be
\frac{\mathrm{D}}{\mathrm{D}t} := \pf{}{t} + \bfv \cdot \pf{}{ \bfx} \ .
\ee
Thermal power is transferred into the system by the heat flux vector $\heatflux$ and generated within the system by the external heat source $\rho\bar{\heatsource}$. Let us define that the rate of the total energy of the body is increased when the heat flux vector is pointing into the body. Then, the thermal power is given by
\bea
\powerT &=& -\intPo \heatflux\cdot\bfn \dd a + \into \rho\,\bar{\heatsource} \dd v = - \into \nablax \cdot \heatflux \dd v + \into \rho\,\bar{\heatsource} \dd v \notag \\
\label{eq:DefQd}
&=& - \intO \nablaX \cdot \heatflux^\refc \dd V + \intO \rho^\refc\,\bar{\heatsource} \dd V
\eea
where $\nablax\cdot\bullet \equiv \mathrm{div}\bullet$ indicates the divergence in the current configuration. \rev{5}{The superscript $\bullet^\comp{0}$ refers to quantities evaluated in the reference configuration and $\rho^\refc= J \, \rho$.} The divergence in the reference configuration is indicated by $\nablaX\cdot\bullet \equiv \mathrm{DIV}\bullet$. Thus, the amount of thermal exchange of the body with its surrounding $\calQ$ is given by time integration of $\powerT$ which yields
\be
\label{eq:DefQ}
\calQ =  -\intO\int\nablaX\cdot\heatflux^\refc \dd t \dd V  + \intO \int \rho^\refc\,\bar{\heatsource} \dd t \dd V \ .
\ee
Inserting the definitions for the internal energy $\calE$, the kinetic energy $\calK$, for the mechanical power $\powerM$ and the thermal power $\powerT$ into the balance of energy \eqref{eq:BalEnergyGlob}, we obtain the local form in the current configuration
\be
\label{eq:BalEnergyLoc}
\rho\,\dot{\bar{\Psi}} + \rho\, \dot{(\theta \overline{s})} = \bfsigma:\bfD - \nablax\cdot\heatflux + \rho\,\bar{\heatsource}
\ee
and in the reference configuration 
\be
\label{eq:BalEnergyLocref}
\rho^\refc \, \dot{\bar{\Psi}} + \rho^\refc\,\dot{(\theta \overline{s})} = \bfP:\dot{\bfF} - \nablaX\cdot\heatflux^\refc + \rho^\refc\,\bar{\heatsource} \ ,
\ee
see \cite{altenbach2012kontinuumsmechanik}, which have to hold for arbitrary processes. Here, the 1$^\text{st}$ Piola-Kirchhoff stress tensor has been used which is related to the Cauchy stress tensor by $\bfP = \rev{8}{\bfsigma\cdot\mathrm{cof}\bfF}$ with the cofactor tensor $\mathrm{cof}\bfF :=  J \, \rev{8}{\bfF^{-T}}$ which transforms vectorial area elements from the reference configuration\rev{8}{, termed as $\mathrm{d}\bfa^\refc$,} to the current configuration\rev{8}{, termed as $\mathrm{d}\bfa$, via $\mathrm{d}\bfa=\mathrm{cof}\bfF\cdot\mathrm{d}\bfa^\refc$}. \rev{7}{Analogously, the heat flux vectors in reference and current configuration are related by $\heatflux^\refc =  \heatflux\cdot\mathrm{cof}\bfF$.} It is worth mentioning that \eqref{eq:BalEnergyLoc} and \eqref{eq:BalEnergyLocref} give the energy balance for thermo-mechanically coupled problems. In case of other coupled problems, i.e. for chemical or electrical coupling, the energy balance has to be modified accordingly. For instance, the formulation of the free energy and the interaction to the surrounding in terms of work or power have to be adapted.

For rigid particles in the isothermal case, the notation
\be
U+K=W
\ee
is usually introduced. Here, $U$ denotes the potential energy, which is associated to the free energy of continuous deformable solids, and $K$ \re{denotes} the kinetic energy. The mechanical work is given by $W=W^\text{in}+W^\text{ex}$ with the contributions \rev{9}{$W^\text{in}$ and $W^\text{ex}$ due to internal and external forces, respectively}.

\subsection{The second law of thermodynamics}
The second law of thermodynamics accounts for the observation that the flow direction of heat is not universal: two bodies of different temperatures will compensate their thermal difference by lowering the temperature of the body with initially higher temperature to increase the temperature of the body with initially lower temperature. This process is irreversible since a ``segregation'' of temperature does not take place.

The unidirectional flow of heat is recast in mathematical terms by the condition that the heat-related energy may only increase. This energy was identified to be the entropy such that Clausius' theorem reads
\be
\Dt{} \into \rho\, \bar{s} \dd v  \ge 0 \ .
\ee
The entropy $s$ may change in time due to entropy flux $\entropyflux$, entropy source $\rho\,\entropysource$, and entropy production $\diss$. It is worth mentioning that precisely the entropy production $\diss$ accounts for dissipative processes which are always path-dependent. Thus, the amount of entropy produced\re{, termed as $\dissI$,} is not given by \re{time integration of the production term $\diss$, i.e.}
\be
\dissI(t_1) - \dissI(t_0) \not= \int_{t_0}^{t_1} \diss \dd t \ .
\ee

The balance of entropy can be written as
\be
\label{eq:BalEntropy0}
\Dt{}\into \rho\,\bar{s} \dd v = -\intPo \entropyflux\cdot\bfn \dd a + \into \rho\,\entropysource \dd v + \into \diss \dd v
\ee
where entropy flux and entropy source are given by $\entropyflux=\heatflux/\theta$ and $\entropysource=\bar{\heatsource}/\theta$, respectively, see e.g. \cite{altenbach2012kontinuumsmechanik}. Then, \eqref{eq:BalEntropy0} locally demands, in the current configuration,
\be
\label{eq:BalEntropy}
\rho\,\dot{\bar{s}} = -\nablax\cdot\Big(\frac{\heatflux}{\theta}\Big) + \rho\,\frac{\bar{\heatsource}}{\theta} + \diss  = \frac{1}{\theta^2}\heatflux\cdot\nablax\theta - \frac{1}{\theta} \nablax\cdot\heatflux + \rho\,\frac{\bar{\heatsource}}{\theta} + \diss
\ee
and
\be
\label{eq:BalEntropyref}
\rho^\refc\,\dot{\bar{s}} = -\nablaX\cdot\Big(\frac{\heatflux^\refc}{\theta}\Big) + \rho^\refc\,\frac{\bar{\heatsource}}{\theta} + \ddDsrefc = \frac{1}{\theta^2} \heatflux^\refc\cdot\nablaX\theta - \frac{1}{\theta} \nablaX\cdot\heatflux^\refc + \rho^\refc\,\frac{\bar{\heatsource}}{\theta} + \ddDsrefc
\ee
in the reference configuration.

The observation that mechanical energy can be transformed completely into heat but the reverse transformation of heat into mechanical energy can only be performed partly, gives rise to define the dissipated energy: it measures the amount of heat that cannot be transformed (back) into mechanical energy, thus into the free energy density. The process of transformation of mechanical energy into non-recoverable thermal energy, which is referred to as dissipation, can be described by restricting $\diss\ge 0$. An idealized, reversible process in which no dissipation takes place is identified by $\diss=0$; a more general case of irreversible processes is identified by $\diss>0$. 

Making use of $\diss\ge0$, we reformulate \eqref{eq:BalEntropy} as follows:
\begin{align}
&& \rho\,\dot{\bar{s}} + \nablax\cdot\entropyflux - \rho\,\entropysource = \diss & \ge 0 \notag \\
\Rightarrow && \rho\,\dot{\bar{s}} + \nablax\cdot\entropyflux - \rho\,\entropysource & \ge 0 \notag \\
\label{eq:ClausiusDuhem}
\Rightarrow && \rho\,\dot{\bar{s}} & \ge - \nablax\cdot\entropyflux + \rho\,\entropysource
\end{align}
where \eqref{eq:ClausiusDuhem} is the well-known Clausius-Duhem inequality \cite{altenbach2012kontinuumsmechanik}.

Combination of the balance of entropy according to \eqref{eq:BalEntropy} and \eqref{eq:BalEntropyref}, respectively, with the energy balance in current or reference configuration in \eqref{eq:BalEnergyLoc} and \eqref{eq:BalEnergyLocref} gives rise to \rev{11}{modified} form of the second law of thermodynamics in the current configuration as
\be
\label{eq:ReducedForm}
\theta \ddDs = \bfsigma:\bfD - \rho\, \dot{\bar\Psi} - \rho\, \dot{\theta} \bar{s} - \frac{1}{\theta} \heatflux\cdot\nablax\theta \ge 0
\ee
and in the reference configuration as
\be
\label{eq:ReducedFormref}
\theta \ddDsrefc = \bfP:\dot{\bfF} - \rho^\refc\, \dot{\bar{\Psi}} - \rho^\refc\, \dot{\theta} \bar{s} - \frac{1}{\theta} \heatflux^\refc \cdot\nablaX \theta \ge 0 \ ,
\ee
respectively. Application of the Coleman-Noll procedure, which is equivalent to the standard argument of dissipative media, results in the constitutive equations for the stresses
\be
\bfP =2 \rho^\refc \,  \bfF \cdot \pf{\bar{\Psi}}{\bfC}
\ee
and entropy
\be
\label{eq:ConstLawEntropy}
\bar{s} = - \pf{\bar{\Psi}}{\theta} \ .
\ee
Furthermore, the relation
\be
-\frac{1}{\theta} \heatflux^\refc\cdot\nablaX\theta \ge 0
\ee
\rev{13}{needs to be satisfied} which can be directly fulfilled by Fourier's law
\be
\heatflux^\refc = - \re{\lambda} \nablaX\theta
\ee
with the heat conductivity \rev{14}{$\lambda>0$}. Finally, the \rev{11}{the second law of thermodynamics reduces to the} so-called dissipation inequality
\be
\label{eq:Biot0}
\theta \dissrefc = \bfp \cdot\dot{\bfupsilon} > 0
\ee
which needs to be satisfied. Here, we used the definition of the thermodynamic driving force $\bfp$ as
\be
\label{eq:DefDrivingForce}
\bfp := - \rho^\refc\,\pf{\bar{\Psi}}{\bfupsilon} \ .
\ee
It is important to mention that $\theta\dissrefc$ is not defined by $\bfp\cdot\dot\bfupsilon$ in \eqref{eq:Biot0}: the entropy production remains an additional quantity that needs to be modeled. Since the case of reversible processes in solids with $\dissrefc=0$ was associated with a non-evolving microstructure, i.e., $\dot{\bfupsilon}=\boldsymbol{0}$, it is obvious that for the present case of irreversible processes the entropy production is related to the evolution of microstructure. Thus, in analogy to the assumption for mechanical power in \eqref{eq:DefWd}, it is convenient to postulate
\be
\label{eq:DefDiss}
\dissrefc = \bfpdissrefc\cdot\dot{\bfupsilon} 
\ee
for the entropy production with the non-conservative forces $\bfpdissrefc$ that have to be specified for a specific material. Provided that $\dot{\bfupsilon}$ is known, it is worth mentioning that the non-conservative forces $\bfpdissrefc$ have to be modeled rather than $\dissrefc$. Consequently, \eqref{eq:Biot0} transforms to the\\
\Graybox{\textbf{Evolution equation for the internal variables $\bfupsilon$}
\be
\label{eq:Biot1}
\bfpdissrefc = \frac{1}{\theta} \bfp \ .
\ee
This equation indicates that the thermodynamic driving force $\bfp$ and the non-conservative force $\bfpdissrefc$ are in balance. It serves as condition to describe the non-conservative evolution of microstructure provided that $\bfpdissrefc$ has been chosen appropriately to the observed material behavior. It is worth mentioning that this local investigation yields an \textbf{ordinary differential equation (ODE)} as evolution equation for $\bfupsilon$.}
\vskip1.0\baselineskip
Note that in case of other couplings than thermo-mechanics the relations have to adapted accordingly.

\section{The principle of stationary action}
Here, the basics of the principle of stationary action for rigid particles is shortly recapitulated to set the basis for its generalization leading to Hamilton's principle. 
\subsection{The action functional for rigid particles}
Routing back to the pioneering works of Maupertius and Euler, a variational approach to modeling the dynamic behavior of physical systems was established by the axiom that the momentum integrated along the path a rigid particle follows tends to be stationary. This can be recast in formulas as
\be
\calA := \int_{\bfu_0}^{\bfu_1} \bfl \cdot \mathrm{d} \bfu \rightarrow \underset{\re{\bfu}}{\text{stat}}
\ee
with the momentum vector
\be
\bfl = m\, \dot{\bfu}
\ee
for a rigid particle with mass $m$ and velocity $\dot{\bfu} = \mathrm{d}\bfu/\mathrm{d} t$. Here, the quantity $\calA$ is referred to as action. Accounting for the time dependence of the spatial coordinate, i.e. $\bfu=\bfu(t)$ such that $\mathrm{d}\bfu=\dot{\bfu}\, \mathrm{d}t$, the action can be reformulated as
\be
\calA = \int_{t_0}^{t_1} m ||\dot{\bfu}||^2 \dd t \ .
\ee
Making use of the kinetic energy
\be
K = \frac{1}{2} m  ||\dot{\bfu}||^2  \ ,
\ee
it can be concluded that the action turns into
\be
\label{eq:Action2K}
\calA =  \int_\tau (K+K) \dd t
\ee
with the arbitrary time interval $\tau=t_1 -t_0$.

\subsection{The principle of stationary action for conservative processes}
The action can be brought to a form that is much more familiar in the context of modern mechanics. To this end, energy conservation as
\be
\int_\tau U \dd t +\int_\tau K \dd t = \re{0} 
\ee
is used to express ``one'' kinetic energy in $\calA=\int_\tau (K+K) \dd t$ in \eqref{eq:Action2K} by $\int_\tau K \mathrm{d} t=
-\int_\tau U \mathrm{d} t$. 
Thus, the action transforms to
\be
\label{eq:ActionRigidParticle}
\calA = \int_\tau (K-U 
) \dd t \ .
\ee
In contrast to the kinetic energy, the potential energy depends on the placement of the rigid particle given by the deformation $\bfu$. The velocity is, per se, not an additional quantity but can be computed from $\bfu$ such that the principle of stationary action is specified to
\be
\calA = \int_\tau L \dd t \rightarrow \underset{\bfu}{\text{stat}}
\ee
with the Lagrange function $L:=K-U$. 
The necessary condition for $\calA$ being stationary is
\be
\delta\calA = \int_\tau \left[ \pf{L}{\dot{\bfu}}\cdot\delta\dot{\bfu} + \pf{L}{\bfu}\cdot\delta\bfu \right] \dd t=  0 \qquad\forall \ \delta\bfu \ .
\ee
Application of integration by parts and postulating $\delta\bfu|_{t_0}=\delta\bfu|_{t_1}=\boldsymbol{0}$  yields the so-called Lagrange equations of 2$^\text{nd}$ kind, reading
\be
\label{eq:PrincipleLeastActionCons}
\frac{\mathrm{d}}{\mathrm{d}t} \pf{L}{\dot{\bfu}} - \pf{L}{\bfu} = \boldsymbol{0} \ ,
\ee
which provides an equivalent formulation of Newton's law
\be
\bff^\text{cons} = m \, \ddot{\bfu} 
\ee
with the conservative forces $\bff^\text{cons}$.

\subsection{The principle of stationary action for non-conservative processes}
In many practical problems, the restriction to conservative particles is an inadmissible simplification of the physical reality. Therefore, non-conservative effects should also be taken into account in the principle of stationary action. The necessary manipulation, however, is small: instead of using energy conservation, the energy theorem 
\be
U+K=W^\text{in} + W^\text{ex}
\ee
is used with the work due to non-conservative internal and external forces, denoted by $W^\text{in}$ and $W^\text{ex}$, respectively. The work due to non-conservative internal forces is given by $W^\text{in}=-\bff^\text{in}\cdot\bfu$ since the non-conservative internal force $\bff^\text{in}$ points in opposite direction to the $\bfu$. The work due to non-conservative external forces $\bff^\text{ex}$ is given by $W^\text{ex}= \bff^\text{ex}\cdot\bfu$. Consequently, the action functional transforms to
\be
\calA = \int_\tau \left(K-U+W^\text{in}+ W^\text{ex}\right) \dd t = \int_\tau \left(L + W^\text{in} + W^\text{ex}\right)  \dd t \rightarrow \underset{\bfu}{\rev{15}{\text{stat}}} \ .
\ee
The variation yields
\be
\delta\calA = \int_\tau \left[ \pf{L}{\dot{\bfu}}\cdot\delta\dot{\bfu} + \pf{L}{\bfu}\cdot\delta\bfu + \pf{W^\text{in}}{\bfu}\cdot\delta\bfu + \pf{W^\text{ex}}{\bfu}\cdot\delta\rev{16}{\bfu}  \right] \dd t = 0 \qquad\forall \ \delta\re{\bfu}
\ee
such that the Euler equation constitutes after integration by parts as
\be
\label{eq:PLA-non-cons}
\frac{\mathrm{d}}{\mathrm{d}t} \pf{L}{\dot{\bfu}} - \pf{L}{\bfu} = \pf{W^\text{in}}{\bfu}  + \pf{W^\text{ex}}{\bfu} 
\ee
when we postulate vanishing variations at the arbitrarily chosen time points $t_0$ and $t_1$ which defined the interval, i.e., $\delta\bfu(t_0) = \delta\bfu(t_1)=\boldsymbol{0}$. The relation \eqref{eq:PLA-non-cons} provides an equivalent formulation of Newton's law
\be
\bff = m \, \ddot{\bfu}
\ee
with the (conservative and non-conservative) resultant forces $\bff$.

\section{Hamilton's principle}
\subsection{The action functional for continuous deformable bodies}
In analogy to rigid particles, the action functional can be assigned for each material point in continuous deformable bodies as
\be
\mathrm{d}\calA := \int_{\bfu_0}^{\bfu_1} \mathrm{d}m^\refc\, \dot{\bfu}\cdot\mathrm{d} \bfu = \int_{\bfu_0}^{\bfu_1} \rho^\refc\, \dot{\bfu}\cdot\mathrm{d} \bfu \  \mathrm{d}V
\ee
such that the action functional of the total body is given by
\be
\calA = \int_\calA \mathrm{d}\calA = \intO \int_{\bfu_0}^{\bfu_1} \rho^\refc\, \dot{\bfu}\cdot\mathrm{d} \bfu \dd V \ .
\ee
In analogy to the treatment of the action for rigid particles, we transform the increment by $\mathrm{d}\bfu=\dot{\bfu}\mathrm{d}t$. This yields
\be
\label{eq:ActionContinuous}
\calA =  \intO \int_\tau \rho^\refc \,  ||\dot{\bfu}||^2 \dd t \dd V   = \int_\tau\intO \rho^\refc \,  ||\dot{\bfu}||^2 \dd V \dd t  = \int_\tau (\calK+\graybox{\calK}\, ) \dd t \ .
\ee

\subsection{The classical Hamilton functional for conservative continua}
Classically, Hamilton's functional equals the action functional $\calA$ when the reduced energy theorem $\calK = - \calE + \calW + \cint$ is inserted into the second term in $\int_\tau (\calK+\calK)\dd t$ in \eqref{eq:ActionContinuous}, yielding
\be
\label{eq:HamiltonFunctionalClassic0}
\calH^\text{classical} := \int_\tau (\calK - \calE + \calW + \cint ) \dd t \ .
\ee
\re{The integration constant $\cint$ does not affect the following results and can be set arbitrarily to zero.} The reduced energy theorem neglects thermal contributions by dropping $\calQ$ such that it is restricted to isothermal processes. Consequently, the free energy in \eqref{eq:DefE} reduces to $\calE = \intO \rho^\refc\bar{\Psi} \dd V $.
Making use of the formula for the mechanical work $\calW$ in \eqref{eq:DefW}, we obtain for Hamilton's functional in its classical sense
\be
\label{eq:calHclassical}
\calH^\text{classical} = \int_\tau \Big( \calK- \intO \rho^\refc\,\bar{\Psi} \dd V + \intPO \bft^{\star\,\refc} \cdot\bfu \dd A + \intO \bfb^{\star\,\refc}\cdot\bfu \dd V  
\Big) \dd t \ .
\ee
We introduce the total potential to be defined as 
\be
\label{eq:Gibbs}
\calG := \intO \rho^\refc\,\bar{\Psi} \dd V - \intPO \bft^{\star\,\refc}\cdot\bfu \dd A - \intO \bfb^{\star\,\refc} \cdot \bfu \dd V \ .
\ee
The first integral $\intO \rho^\refc\,\bar{\Psi} \dd V=:\Pi^\text{in}$ is the internal potential which depends on the current physical state expressed in terms of distortions, (constant) temperature and the (frozen) microstructure. The last two integrals $- \intPO \bft^{\star\,\refc}\cdot\bfu \dd A - \intO \bfb^{\star\,\refc} \cdot \bfu \dd V=:\Pi^\text{ex}$ represent the potential due to external forces which depends on the current boundary conditions. Consequently, by inserting \eqref{eq:Gibbs} into \eqref{eq:calHclassical}, we obtain
\be
\label{eq:calHclassical1}
\calH^\text{classical} = \int_\tau \Big(\calK - \calG  
\Big) \dd t \ .
\ee
Assuming stationarity for $\calH^\text{classical}$ with respect to the only remaining variable, which are the displacements $\bfu$, the integration constant $\cint$ does not have any influence and can be set to zero. Thus, Hamilton's functional corresponds to the representation of the action in \eqref{eq:ActionRigidParticle} for conservative continuous deformable bodies. Note the similarity to the principle of the minimum of the total potential energy: however, whereas the Hamilton principle only demands stationarity of \eqref{eq:calHclassical1}, the principle of minimum potential energy requires minimizers. In practice, this leads to the same conditions as only the first variation is analyzed for both cases. Making use of the potentials $\Pi^\text{in}$ and $\Pi^{\text{ex}}$, \eqref{eq:calHclassical1} can be represented by
\be
\calH^\text{classical} = \int_\tau \Big( \calK - \Pi^\text{in} - \Pi^\text{ex} \Big) \dd t \ .
\ee

\subsection{The classical Hamilton principle}
Comparably to the axiom of stationary action for rigid particles, Hamilton's principle in its classical sense reads
\be
\calH^\text{classical} \rightarrow \underset{\bfu}{\text{stat}} \qquad \Leftrightarrow \qquad \delta_{\bfu}\calH^\text{classical} = 0 \quad\forall \ \delta\bfu \ .
\ee
Computation of the G\^ateaux derivative results in
\be
\int_\tau \Big( \intO \rho^\refc\,\dot{\bfu}\cdot\delta\dot{\bfu} \dd V - \intO \rho^\refc \pf{\bar{\Psi}}{\bfC} :\delta\bfC \dd V + \intPO \bft^{\star\,\refc}\cdot\delta\bfu \dd A - \intO \bfb^{\star\,\refc} \cdot \delta\bfu \dd V \Big) \dd t = 0
\ee
and after integration by parts
\be
-\intO \rho^\refc \, \ddot{\bfu}\cdot\delta\bfu \dd V + \intO \left(\nablaX\cdot\re{\bfP^\text{T}} + \bfb^{\star\,\refc} \right) \cdot\delta\bfu \dd V + \intPO \left(\bfn^\refc\cdot\re{\bfP^\text{T}} - \bft^{\star\,\refc}\right)\cdot\delta\bfu \dd A = 0  
\ee
which yields the \rev{18}{balance of} linear momentum and Cauchy's theorem \rev{18}{as boundary condition}. Consequently, classical Hamilton's principle can be interpreted as a generalization of the principle of stationary action. Conversely, the principle of stationary action constitutes a special case of the classical Hamilton principle when deformations are neglected (rigid bodies).

\subsection{The extended Hamilton functional for non-conservative continua}
The restriction to isothermal processes of Hamilton's functional in its classical form according to \eqref{eq:calHclassical1} also prevents to account directly for dissipation due to microstructure evolution.
Consequently, we propose to replace one $\calK$ by the entire energy, thus $\calK = - \calE + \calW + \calQ + \cint$.  This yields
\be
\calH = \int_\tau \left(\calK - \calE + \calW + \graybox{\calQ} + \cint \right) \dd t
\ee
where consideration of the thermal work $\calQ$ indicates the extension of the classical Hamilton functional. For other coupled problems, the replacement of one kinetic energy term has to be adapted according to the respective form of energy conservation, i.e. taking into account Poynting's theorem which includes Joule heating \cite{poynting1884xv}, for instance.
Inserting the formulas for internal energy $\calE$, mechanical work $\calW$ and thermal work $\calQ$ introduced in Eqs. \eqref{eq:DefE}, \eqref{eq:DefW}, and \eqref{eq:DefQ}, \re{and setting $\cint=0$,} we obtain for Hamilton's functional
\bea
\calH &=& \int_\tau \Big( \calK- \intO \rho^\refc\,\bar{\Psi} \dd V - \intO \rho^\refc\,\theta \bar{s} \dd V + \intPO \bft^{\star\,\refc} \cdot\bfu \dd A  \notag \\
\label{eq:calH00}
&& \quad + \intO \bfb^{\star\,\refc}\cdot\bfu \dd V  - \intPO \int\heatflux^\refc\cdot\bfn^\refc \dd t \dd A + \intO \int \rho^\refc\,\bar{\heatsource} \dd t \dd V 
\Big) \dd t \ .
\eea
By inserting \eqref{eq:Gibbs} into \eqref{eq:calH00}, we receive for Hamilton's functional
\be
\label{eq:Hamilton5}
\calH = \int_\tau \Big(\calK - \calG - \intO \rho^\refc\,\theta \bar{s} \dd V - \intPO \int \heatflux^\refc\cdot\bfn^\refc \dd t\dd A + \intO \int \rho^\refc\,\bar{\heatsource} \dd t \dd V 
\Big) \dd t \ .
\ee
In contrast to the classical form of Hamilton's function according to \eqref{eq:calHclassical1}, the extended form in \eqref{eq:Hamilton5} also accounts for non-isothermal contributions which covers both conservative and non-conservative processes that are accompanied by changes of the state variable $\theta$. This extension, as we will show in the following, allows to derive field equations for all physical state variables including temperature and internal variables.

For the following derivations, it is beneficial to reformulate $\calH$. To this end, let us make use of the entropy equivalence according to \eqref{eq:BalEntropyref} as
\be
\rho^\refc\, \theta \dot{\bar{s}}  = \frac{1}{\theta} \heatflux^\refc\cdot\nablaX\theta -\nablaX\cdot\heatflux^\refc  + \rho^\refc\,\bar{\heatsource} +  \theta\ddDsrefc  \ .
\ee
Integration by parts yields
\begin{align}
 && \int \rho^\refc\, \theta \dot{\bar{s}} \dd t &=  \rho^\refc\, \theta \bar{s} - \int \rho^\refc\, \dot{\theta}\bar{s} \dd t \notag \\
 &&& =  \int \frac{1}{\theta} \heatflux^\refc\cdot\nablaX\theta \dd t -\int\nablaX\cdot\heatflux^\refc \dd t  + \int \rho^\refc\,\bar{\heatsource} \dd t + \int \theta\ddDsrefc \dd t \notag \\
\label{eq:xx1}
\Leftrightarrow && -\rho^\refc\,\theta \bar{s} - \int \nablaX\cdot \heatflux^\refc \dd t & + \int \rho^\refc\,\bar{\heatsource} \dd t  = - \int \rho^\refc\,\dot{\theta}\bar{s} \dd t - \int  \frac{1}{\theta} \heatflux^\refc\cdot\nablaX\theta \dd t - \int \theta\ddDsrefc \dd t
\end{align}
when mass conservation $\dot{\rho}^\refc=0$ is considered. As already mentioned above, the antiderivative $\int\theta\ddDsrefc\, \mathrm{d}t$ does not exist as rational ma\-the\-ma\-ti\-cal formula: application of the fundamental theorem of calculus would suppress the path-dependence of the entropy production. Consequently, the anitderivative has to be modeled appropriately if necessary. 
Finally, inserting \eqref{eq:xx1} into \eqref{eq:Hamilton5} yields the\\
\Graybox{\textbf{Extended Hamilton functional}
\be
\label{eq:HamiltonFunctionalNonConsCont}
\calH = \int_\tau \Big(\calK -\calG - \intO \int\rho^\refc\,\dot{\theta} \bar{s} \dd t \dd V - \int_\Omega \int  \frac{1}{\theta} \heatflux^\refc\cdot\nablaX\theta \dd t \dd V -  \intO \int\theta\ddDsrefc \dd t \dd V 
\Big) \dd t 
\ee
which is valid for thermo-mechanically coupled problems and dissipative microstructure evolution.}
\vskip\baselineskip
It is worth mentioning that the presentation of Hamilton's functional in \eqref{eq:HamiltonFunctionalNonConsCont} includes antiderivatives with respect to time, i.e. $\int \bullet \dd t$, which constitutes a generalization of Hamilton's functional in the classical sense, where only isothermal processes have been accounted for. Consequently, the unspecific time integrals are not present in the classical form. However, they resemble to methods using time-incremental variations, see e.g. \cite{simo1990variational,miehe2002strain,miehe2011multi}.

\subsection{The extended Hamilton principle for non-conservative continua}
The extended Hamilton principle for non-conservative continua postulates that Hamilton's functional in \eqref{eq:HamiltonFunctionalNonConsCont} tends to be stationary for the state variables. Usually, the stationarity conditions with respect to displacements and internal variables are investigated. However, we propose that the stationarity conditions with respect to all state variables shall be investigated. This, obviously,  comprises the stationarity condition with respect to temperature. Then, the stationarity is recast in formulas as
\be
\calH \rightarrow \underset{\bfu,\theta,\bfupsilon}{\text{stat}} \qquad \Leftrightarrow \qquad \delta\calH = \delta_{\bfu} \calH + \delta_\theta \calH + \delta_{\bfupsilon} \calH = 0 \quad \forall \ \delta\bfu, \, \delta\theta, \, \delta\bfupsilon \ .
\ee
Since the variations of $\bfu$, $\theta$, and $\bfupsilon$ are independent, the necessary conditions for stationarity constitute as 
\be
\begin{cases}
\delta_{\bfu}\calH = 0 & \forall \ \delta\bfu \\[2mm]
\delta_\theta\calH = 0 & \forall \ \delta\theta\\[2mm]
\delta_{\bfupsilon}\calH = 0 & \forall \ \delta\bfupsilon
\end{cases} \ .
\ee
This set of equations describes, in a unified way, the evolution of the thermomechanical state of solids, given in terms of $\{\bfu,\theta,\bfupsilon\}$. Let us evaluate the stationarity conditions by computing the G\^{a}teaux derivatives with respect to $\bfu$, $\theta$, and $\bfupsilon$. They read
\be
\label{eq:GateauxAll}
\begin{cases}
\D \delta_{\bfu} \calH = \int_\tau \Big(\delta_{\bfu} \calK - \delta_{\bfu} \calG \Big) \dd t = 0 & \forall \ \delta\bfu \ , \\[4mm]
\D \delta_\theta \calH = \int_\tau \Big(  - \delta_\theta \calG - \intO \int \delta_\theta (\rho^\refc\,\dot{\theta}\bar{s}) \dd t \dd V \\[4mm]
\D\qquad\quad - \int_\Omega \int \delta_\theta \Big(\frac{1}{\theta} \heatflux^\refc\cdot\nablaX\theta\Big) \dd t \dd V - \intO \int \delta_\theta \big(\theta\ddDsrefc\big) \dd t \dd V \Big) \dd t = 0 & \forall \ \delta\theta \ , \\[4mm]
\D \delta_{\bfupsilon} \calH = \int_\tau \Big( -\delta_{\bfupsilon} \calG - \intO \int \delta_{\bfupsilon} \big(\theta\ddDsrefc\big) \dd t \dd V \Big) \dd t = 0 & \forall \ \delta\bfupsilon \ .
\end{cases}
\ee
We postpone the detailed evaluation of \eqref{eq:GateauxAll} to Sec.~\ref{sec:Derivations} and present directly the results, which are the\\
\Graybox{\textbf{Field equations for the thermo-mechanical state variables $\bfu,\theta,\bfupsilon$}
\bea
\label{eq:FieldEquationU}
&&\begin{cases} \D\intO \left(\nablaX\cdot\rev{19}{\bfP^\text{T}} + \bfb^{\star\,\refc} \right) \cdot\delta\bfu \dd V = \intO \rho^\refc \, \ddot{\bfu}\cdot\delta\bfu \dd V\\[3mm]
\D \intPO \left(\bfn^\refc\cdot\rev{19}{\bfP^\text{T}} - \bft^{\star\,\refc}\right)\cdot\delta\bfu \dd A = 0  
\end{cases} \\[3mm]
\label{eq:FieldEquationT}
&&\begin{cases}
\D \intO  \left(\kappa \ \dot{\theta} \re{+} \nablaX\cdot\heatflux^\refc + \rev{28}{\left[\pf{\Psi^\refc}{\bfupsilon} + \theta \, \pf{}{\theta}\pf{\Psi^\refc}{\bfupsilon} \right] \cdot\dot{\bfupsilon} - 2 \theta \, \pf{\bfS}{\theta} :\dot{\bfC} } \right) \delta\theta  \dd V = 0 \\[3mm]
\D \intPO  \bfn^\refc\cdot \heatflux^\refc \ \delta \theta  \dd A = 0
\end{cases}  \\[3mm]
&& \begin{cases}
\label{eq:FieldEquationA}
\D\intO \Big( \pf{\Psi^\refc}{\bfupsilon} - \nablaX\cdot \pf{\Psi^\refc}{\nablaX\bfupsilon} + \pf{\Pidissrefc}{\dot{\bfupsilon}} \Big)\cdot\delta\bfupsilon \dd V = 0 \\[3mm]
\D \intPO \bfn^\refc\cdot \pf{\Psi^\refc}{\nablaX\bfupsilon} \cdot \delta\bfupsilon \dd A  = 0 
\end{cases}  
\eea
It is worth mentioning that the balance of linear momentum in \eqref{eq:FieldEquationU} results from Hamilton's principle in its classical form, which is equivalent to the stationarity of the total potential in the quasi-static case, i.e.~$\ddot{\bfu}\approx\boldsymbol{0}$.~However, the stationarity of the extended Hamilton functional yields also the heat equation in~\eqref{eq:FieldEquationT}. \rev{28}{Here, two heat sources are present without any further assumptions: i) the heat production due to microstructure evolution; and ii) the thermo-elastic coupling term $2\theta\pf{\bfS}{\theta}:\dot{\bfC}$. Note that the classical form of the heat conduction equation results if the dependence on temperature of both, the thermodynamic driving force and the 2${}^\text{nd}$ Piola Kirchhoff stress tensor can be neglected.} Furthermore, the stationarity condition with respect to the internal variables in \eqref{eq:FieldEquationA} constitutes a \textbf{partial differential equation} (PDE) which accounts both for local and non-local effects and, thus, is a more general form than the ODE as the evolution equation in \eqref{eq:Biot1}. The system of equations is closed when
\begin{itemize}
\item the free energy density \rev{20}{per undeformed unit volume} $\Psi^\refc$, 
\item the heat flux vector $\heatflux^\refc$, 
\item and the dissipation function $\Pidissrefc$ 
\end{itemize}
have been modeled by suitable constitutive equations. 
}

\vskip1.0\baselineskip
Let us investigate the consequences of the field equations for the thermo-mechanical state variables. 
\subsubsection{Displacements}
We start with \eqref{eq:FieldEquationU}. The first line is identified as the principle of virtual work that is widely used in mechanics and serves as fundamental equation for defining boundary value problems in solid mechanics; furthermore, the Euler-Lagrange equation as
\be
\nablaX\cdot\rev{19}{\bfP^\text{T}} + \bfb^{\star\,\refc} =  \rho^\refc \, \ddot{\bfu} \qquad\forall \ \bfx\in\Omega
\ee
can be identified as the local form of the balance of linear momentum in the reference configuration. The second line is identified as Cauchy's theorem
\be
\rev{21}{\bfn^\refc \cdot \bfP^\text{T} =}\bfP\cdot\bfn^\refc = \bfF\cdot\bfS\cdot\bfn^\refc = \bft^{\star\,\refc} \ \qquad \forall \ \bfx\in\partial\Omega
\ee
which serves as Neumann boundary condition for the displacements. \re{The 2${}^\text{nd}$ Piola Kirchhoff stress tensor is denoted by $\bfS=2\rho^\refc\,\partial\bar{\Psi}/\partial\bfC$.}

\subsubsection{Temperature}
If we use, for instance, Fourier's law $\dot{\bfq}^\refc=-\lambda \nablaX\theta$ with the thermal conductivity $\lambda$ and make use of the definition for the thermodynamic driving forces $\bfp=- \rho^\refc\partial\rev{23}{\bar{\Psi}}^\refc/\partial\bfupsilon=:-\partial\Psi^\refc/\partial\bfupsilon$ according to \eqref{eq:DefDrivingForce}, we thus obtain for \eqref{eq:FieldEquationT}
\be
\kappa \dot{\theta} = \bfp \cdot\dot{\bfupsilon} \re{+} \lambda \nablaX\cdot\nablaX\theta \qquad \forall \ \bfx\in\Omega 
\ee
\re{when we neglect thermo-elastic coupling and a possible dependence of the driving forces on temperature, i.e.~$\pf{\bfS}{\theta}:\dot{\bfC}\approx0$ and~$\pf{\bfp}{\theta}\approx\boldsymbol{0}$. This field equation} is nothing but the heat equation: temperature evolves in time and space while thermodynamic driving force times thermodynamic flux, i.e., $\bfp \cdot\dot{\bfupsilon}$, serve as local process-specific heat source due to microstructural evolution. 

The boundary condition
\be
\bfn^\refc\cdot\nablaX\theta = 0 \qquad \forall \ \bfx\in\partial\Omega
\ee
of vanishing temperature gradient at the surface corresponds to an adiabatic system. Other conditions, i.e. non-vanishing Neumann boundary conditions can be easily considered: a non-zero heat flux at the surface can be accounted for by replacing
\be
\calH \rightarrow \calH^\text{thermal BC} \qquad \text{with}\qquad\calH^\text{thermal BC} = \calH + \int_\tau \intPO \int \tilde{\gamma}^{\star\,\refc} \theta \dd t \dd A \dd t 
\ee
and the temperature-specific heat at the surface $\tilde\gamma^{\star\,\refc}$. Then, $\delta_\theta\calH^\text{thermal BC} =0$ comprises the modified surface condition
\be
 \intPO \int \Big(\theta \ \tilde{\gamma}^{\star\,\refc} - \lambda \bfn^\refc\cdot \nablaX\theta \Big) \delta \theta \dd t \dd A = 0
\ee
for inhomogeneous thermal boundary conditions and consequently
\be
\lambda \bfn^\refc\cdot \nablaX\theta = \gamma^{\star\,\refc} \ .
\ee
The heat at the surface  $\gamma^{\star\,\refc}$ is given in terms of the temperature-specific heat $\tilde{\gamma}^{\star\,\refc}$ through $\gamma^{\star\,\refc} = \tilde{\gamma}^{\star\,\refc} \theta$. Similarly, the extension
\be
\calH \rightarrow \calH^\text{inner heat source} \qquad\text{with}\qquad \calH^\text{inner heat source} = \calH - \int_\tau \intO \int \rho^\refc\,\tilde{h}^{\star\,\refc} \theta \dd t \dd A \dd t
\ee
results in the heat equation
\be
\kappa \dot{\theta} = \bfp \cdot\dot{\bfupsilon} \re{+} \lambda \nablaX\cdot\nablaX\theta + \rho^\refc\,\bar{\heatsource}^{\star}
\ee
with the heat source $\bar{\heatsource}^{\star}=\theta \, \tilde{\heatsource}^{\star\,\refc}$. For isothermal processes, this equation reduces correctly to, cf. \eqref{eq:sisotherm} \re{and \eqref{eq:StatHSimpleMaterial}},
\be
\label{eq:EntropyIsothermal}
\rho^\refc\, \theta \bar{s} = \int \bfpdissTrefc\cdot\dot{\bfupsilon} \dd t \re{= \int \bfp \cdot\dot{\bfupsilon} \dd t = \int -\pf{\Psi}{\bfalpha} \cdot \dot{\bfalpha} \dd t =: \int - \dot{\Psi}_\alpha \dd t}
\ee
\re{with the dissipative part of the rate of the free energy density $\dot{\Psi}_\alpha$.}

\subsubsection{Internal variables}
From \eqref{eq:FieldEquationA}, we find the strong form of the field equation for the internal variables, given as
\be
\label{eq:EvolutionEquationField}
\pf{\Pidissrefc}{\dot{\bfupsilon}} = \bfp + \nablaX\cdot \pf{\Psi^\refc}{\nablaX\bfupsilon} \qquad \forall \ \bfx\in\Omega \ .
\ee
It is worth mentioning that \eqref{eq:EvolutionEquationField} constitutes as a PDE for the internal variables. To this end, the dissipation function $\Pidissrefc$ needs to be modeled. More details are given in Sec.~\ref{ssec:SecificationNCForce}. The Neumann boundary condition is given as
\be
 \bfn^\refc\,\cdot \pf{\Psi^\refc}{\nablaX\bfupsilon} = \rev{24}{\boldsymbol{0}} \qquad \forall \ \bfx\in\partial\Omega \ .
\ee

We present the relation of the stationarity of the extended Hamilton principle to the principle of the minimum of the dissipation potential in Sec.~\ref{ssec:SimpleMaterials} and give the most important modeling approaches for $\Pidissrefc$ in Sec.~\ref{ssec:SecificationNCForce}. Finally, in Sec.~\ref{ssec:FinalRemarks}, we show that material models obtained by the extended Hamilton principle are thermodynamically sound by construction and present the strong form of the stationarity conditions for the case of linearized kinematics.

\subsubsection{Simple materials}
\label{ssec:SimpleMaterials}
 Let us investigate the consequences of the stationarity with respect to the internal variables in more detail. Thus, we start by considering a so-called simple material, i.e. $\Psi^\refc\not=\Psi^\refc(\nablaX\bfupsilon)\Rightarrow\partial\Psi^\refc/\partial\nablaX\bfupsilon=\boldsymbol{0}$, the evolution equation reduces to
 \be
\label{eq:Biot2}
\pf{\calD^{\text{diss}}}{\dot{\bfupsilon}} + \pf{\Psi^\refc}{\bfupsilon} = \boldsymbol{0}
\ee
which we can write in an alternative form as
\be
\label{eq:StatHSimpleMaterial}
\bfpdissTrefc = \bfp
\ee
with the non-conservative force $\bfpdissTrefc=\partial\Pidissrefc/\partial\dot{\bfupsilon}$, cf. \eqref{eq:DefDissiaptedEnergy}, and $\bfp=-\partial\Psi^\refc/\partial\bfupsilon$. We recognize that \eqref{eq:StatHSimpleMaterial} is the well-known Biot's equation \cite{biot1965mechanics,nguyen2000stability} which was postulated as fundamental equation to describe microstructure evolution. Interestingly, assuming stationarity of Hamilton's functional with respect to the internal variables results in the very same equation. From \eqref{eq:Biot1}, we furthermore obtain the relation
\be
\label{eq:NonConsForces}
\bfpdissrefc = \frac{1}{\theta} \bfpdissTrefc
\ee
which indicates that $\bfpdissrefc$ is the temperature-specific non-conservative force.

After integration with respect to $\dot{\bfupsilon}$, a potential form of \eqref{eq:Biot2} reads 
\be
\label{eq:MinDisPot}
\calL := \dot{\Psi}^\refc + \calD^{\text{diss}} \rightarrow \underset{\dot{\bfupsilon}}{\text{stat}}
\ee
when the integration constant is set to $\pf{\Psi^\refc}{\bfC}:\dot{\bfC}+\pf{\Psi^\refc}{\theta}\dot{\theta}$. Consequently, for simple materials, the extended Hamilton principle demands stationarity of the so-called dissipation potential $\calL$ which is known as principle of the minimum of the dissipation potential \cite{hackl1997generalized}. It is worth mentioning that this principle could be extended to non-isothermal processes \cite{junker2013principle}, but there, the heat equation was -- as usual -- derived from the balance of energy. In case of using Hamilton's principle, in contrast, the heat equation follows as stationarity condition of a variational problem.

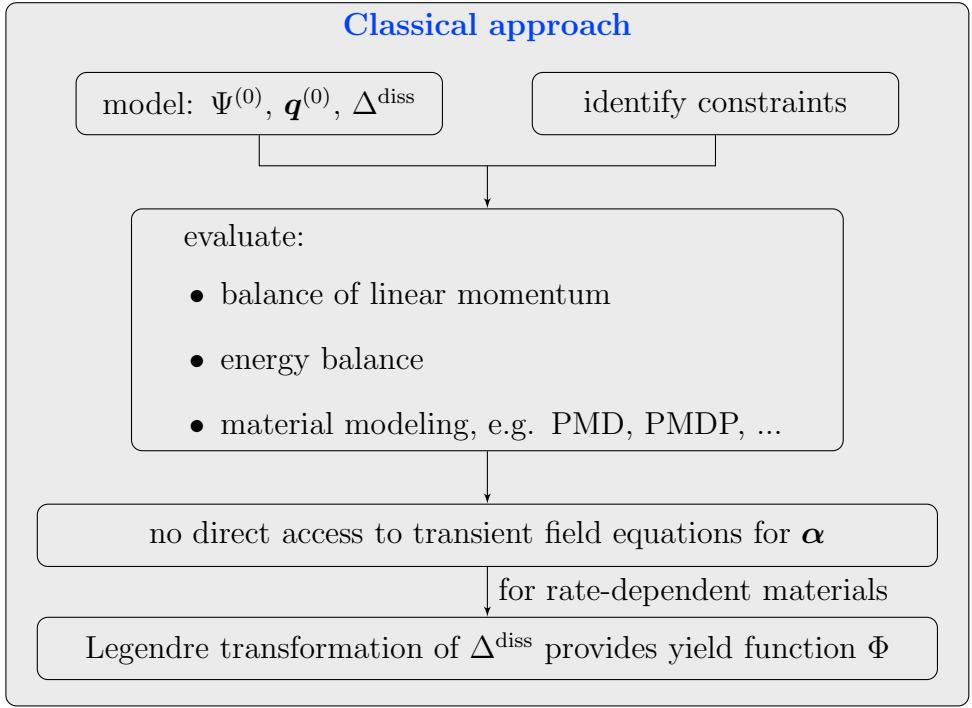
\begin{figure}[hptb]
\centering
    \tikzstyle{blind} = [circle, 
    text width=0em, text centered, minimum height=0em]
\tikzstyle{line} = [draw, -latex']
\tikzstyle{ini1} = [rectangle, draw, 
    text width=30em, text centered, rounded corners, minimum height=2em,fill=lightgray]
\tikzstyle{ini2} = [rectangle, draw, 
    text width=11em, text centered, rounded corners, minimum height=2em,fill=lightgray]
\tikzstyle{ini3} = [rectangle, draw, 
    text width=17em,   text centered, rounded corners, minimum height=2em,fill=lightgray]
\tikzstyle{ini4} = [rectangle, draw, 
    text width=28em,   text centered, rounded corners, minimum height=2em,fill=lightgray]        
\tikzstyle{ini5} = [rectangle, draw, 
    text width=22em,   text centered, rounded corners, minimum height=2em,fill=lightgray]     
    \begin{tikzpicture}[node distance = 2.5cm, auto]
    \node [ini1] (Classical) at (0,-8.0) {\textbf{\textcolor{blau}{Classical approach}}\vspace{8.675cm}   
	};
    \node[ini2](Cmodel) at (-3,-4.675) { 
    model: $\Psi^\refc$, $\heatflux^\refc$, $\Pidissrefc$
    };
    \node[ini2](Ccons) at (3,-4.675) { 
   identify constraints
    };    
    \node[ini5](Cevaluate) at (0,-7.675) {
   \begin{flushleft} \vspace{-5mm} ${}\quad{}$ evaluate: \vspace{-2mm} \end{flushleft}
    \begin{itemize}
    \item balance of linear momentum
    \item energy balance
    \item material modeling, e.g. PMD, PMDP, ...
    \end{itemize}
    };
    \node[ini4](Cresult) at (0,-10.4) {
  no direct access to transient field equations for $\bfupsilon$
    };
    \node[ini4](Cyield) at (0,-11.9) {
  Legendre transformation of $\Pidissrefc$ provides yield function $\Phi$
    };
	\draw  (Cmodel) -- (-3,-5.5) -- (3,-5.5) -- (Ccons);    
	\draw[line] (0,-5.5) -- (Cevaluate);
	\draw[line] (Cevaluate) -- (Cresult);
	\draw[line] (Cresult) -- (Cyield) node[midway] (TextNode2) {for rate-dependent materials};
    \end{tikzpicture}
\caption{
General use of classical approaches: the free energy density $\Psi^\refc$, the heat flux vector $\heatflux^\refc$, and the dissipation function $\Pidissrefc$, sometimes also called dissipation potential or, \rev{30}{although being mathematically imprecise}, dissipation functional, have to be specified along with possible constraints \rev{b}{(instead of modeling $\heatflux^\refc$, appropriate extensions of $\Pidissrefc$ are also possible, cf.~\cite{hackl2011study})}. Alternatively to $\Pidissrefc$, the modeling of the yield function $\Phi$ is possible. Then, balance of linear momentum, which constitutes as strong form of the the minimum of the total potential for quasi-static processes, i.e. $\ddot{\bfu}\approx\boldsymbol{0}$, and the energy balance, yielding the strong form for the heat conduction equation, have to be evaluated. Furthermore, material modeling has to be performed. Possible strategies for the derivation of material models are the principle of maximum dissipation (PMD), the principle of the minimum of the dissipation potential (PMDP), and others. The entire procedure, however, does not provide (direct) access to transient field equations for the internal variables $\bfupsilon$. Furthermore, coupling is not inherently ensured. 
}
\label{fig:FlowC}
\end{figure}

\subsubsection{Specification of the dissipation function and the non-conservative force}
\label{ssec:SecificationNCForce}
Along with the free energy density $\Psi^\refc$ and the heat flux vector $\heatflux^\refc$, the dissipation function $\Pidissrefc$, or alternatively the non-conservative force $\bfpdissTrefc$, has to be defined to close the system of equations. Here, two approaches cover the majority of material behavior:
\be
\bfpdissTrefc = \bfpdissTrefc_\text{RD} := \eta\, \dot{\bfupsilon} \qquad \text{and} \qquad \bfpdissTrefc = \bfpdissTrefc_\text{RI} := r \frac{\dot{\bfupsilon}}{||\dot{\bfupsilon}||} \ .
\ee
Rate-dependent microstructure evolution is covered by $\bfpdissTrefc_\text{RD}$ where $\eta$ is the viscosity. \rev{26}{Hereby, viscous microstructure evolution can be modeled as, for instance, in visco-elasticity.} In contrast, rate-independent microstructure evolution is covered by $\bfpdissTrefc_\text{RI}$ \rev{26}{as it appears in modeling of, e.g., elasto-plastic material behavior}. The related dissipation functions are
\be
\calD^{\text{diss}}_\text{RD} = \frac{1}{2}\eta\, ||\dot{\bfupsilon}||^2 \qquad \text{and} \qquad \calD^{\text{diss}}_\text{RI} = r ||\dot{\bfupsilon}|| \ .
\ee
By using the definition \eqref{eq:DefDiss} and the relation \eqref{eq:NonConsForces} which gives $\dissrefc=\bfpdissrefc\cdot\dot{\bfupsilon}=\bfpdissTrefc\cdot\dot{\bfupsilon}/\theta$, we can relate the dissipation functions to the entropy production as
\be
\ddDsrefcRD = \frac{2}{\theta} \calD^{\text{diss}}_\text{RD} \qquad\text{and}\qquad \ddDsrefcRI = \frac{1}{\theta} \calD^{\text{diss}}_\text{RI}\ .
\ee

In case of rate-independent evolution, an associated indicator function, i.e., a yield function, is missing. Again, two variants are possible. One variant is a so-called non-associated material law, i.e.
\be
\dot{\bfupsilon} \not\propto \pf{\Phi}{\bfp} \ .
\ee
In this case, the \re{yield} function has to be defined independently. The other variant is an associated material law. Here, the \rev{25}{yield} function is not modeled but it results from modeling the dissipation function $\Pidissrefc$. It can be identified by mathematically analyzing the Legendre transformation of $\calD^{\text{diss}}_\text{RI}$ which exchanges the thermodynamic flux $\dot{\bfupsilon}$ by the total thermodynamic driving force $\bfp^\text{total}:= \bfp+\nablaX\cdot\partial\Psi/\partial\nablaX\bfupsilon$. Thus, making use of the dissipation function for rate-independent evolution, \eqref{eq:FieldEquationA} transforms to 
\be
\label{eq:EvolutionEquationPTotal}
r \frac{\dot{\bfupsilon}}{||\dot{\bfupsilon}||} \ni \bfp^\text{total\,\refc} \qquad \Leftrightarrow \qquad \dot{\bfupsilon} \ni \frac{||\dot{\bfupsilon}||}{r} \bfp^\text{total\,\refc}\ .
\ee
Equation \eqref{eq:Biot2} indicates that the derivative of the dissipation function equals the thermodynamic driving force. This allows to apply a Legendre transformation to exchange the rate of internal variables $\dot{\bfupsilon}$ by its conjugated variable which is given by $\bfp^\text{total}$. This Legendre transformation yields \rev{25}{the indicator function}
\bea
\calD_\text{RI}^\text{diss LT} &=& \underset{\dot\bfupsilon}{\text{sup}} \Big\{ \bfp^\text{total}\cdot\dot{\bfupsilon} - \calD^{\text{diss}}_\text{RI} \Big\} \notag \\ 
&=& \underset{\dot\bfupsilon}{\text{sup}} \left\{ \frac{||\dot{\bfupsilon}||}{r} \Big( ||\bfp^\text{total}||^2 - r^2 \Big) \right\}  = \begin{cases} \infty & \text{for } ||\bfp^\text{total}|| > r \\
0 & \text{for } ||\bfp^\text{total}|| \le r \end{cases} 
\eea
when \eqref{eq:EvolutionEquationPTotal} has been inserted. Hence, the \rev{25}{yield function, also known, e.g., from plasticity}, is identified as
\be
\Phi := ||\bfp^\text{total}|| - r \le 0 \ .
\ee
Consequently, the parameter $r$ serves as energetic threshold value for microstructural evolution. 

Combination of the rate-dependent and rate-independent non-conservative forces by
\be
\bfpdissTrefc = \bfpdissTrefc_\text{VP} := \eta\, \dot{\bfupsilon} +  r \frac{\dot{\bfupsilon}}{||\dot{\bfupsilon}||}
\ee
with the related dissipation function
\be
\calD^{\text{diss}}_\text{VP} = \frac{1}{2}\eta\, ||\dot{\bfupsilon}||^2  + r ||\dot{\bfupsilon}||
\ee
results in a visco-plastic material behavior\rev{26}{, i.e., a viscous microstructure evolution takes place once the energetic threshold $r$ has been reached}. The evolution equation resulting form Hamilton's principle in this case reads
\be
\dot{\bfupsilon} = \frac{1}{\eta} \, \Phi_+ \ \frac{\bfp^\text{total}}{||\bfp^\text{total}||}
\ee
where $\Phi_+ := (\Phi + |\Phi|)/2$ denotes the Macaulay bracket.

\subsubsection{Final remarks}
\label{ssec:FinalRemarks}
For the presented approaches for the non-conservative forces, Onsager's principle of proportionality between thermodynamic fluxes and thermodynamic forces is obtained. It is thus obvious that for all cases, rate-dependent, rate-independent, and visco-plastic microstructure evolution, thermodynamic consistency is fulfilled: by inserting the respective evolution equations into the dissipation inequality \eqref{eq:Biot0}, we find for all cases
\be
g\, \bfp^\text{total} \cdot \bfp^\text{total} = g\, ||\bfp^\text{total\,\refc}||^2 \ge 0 
\ee
where $g\ge 0$ since
\be
g=\begin{cases}
\D \eta^{-1}& \text{for rate-dependent processes} \\[2mm]
\D r^{-1} ||\dot{\bfupsilon}|| & \text{for rate-independent processes} \\[2mm]
\D \eta^{-1}\Phi_+||\bfp^{\text{total}\,\refc}||^{-1}  & \text{for visco-plastic material behavior}
\end{cases} \ .
\ee
Consequently, material models derived using Hamilton's principle agree by construction with the fundamental thermodynamic laws if appropriate modeling approaches for the dissipation function $\Pidissrefc$, i.e. or alternatively for non-conservative force $\bfpdissT$, are made.

The local form of the stationarity conditions for the case of linearized kinematics is given by
\be
\begin{cases}
\D \nabla\cdot\bfsigma + \bfb^\star =  \rho \, \ddot{\bfu} \\[4mm]
\D  \kappa\, \dot{\theta} = \rev{28}{\theta \pf{\bfsigma}{\theta}:\dot{\bfeps} + \left[\bfp + \theta \, \pf{\bfp}{\theta}\right] \cdot\dot{\bfupsilon} } \re{+} \lambda \nabla\cdot\nabla\theta \\[4mm]
\D \pf{\Pidissrefc}{\dot{\bfupsilon}} = \bfp + \rev{27}{\nabla\cdot\pf{\Psi}{\nabla\bfalpha}}
\end{cases} \qquad \forall\ \bfx\in\Omega
\ee
with the constitutive relation $\bfsigma=\partial\Psi/\partial\bfvarepsilon$, the strain tensor $\bfvarepsilon=(\nabla\bfu+\bfu\nabla)/2$, the heat flux vector $\heatflux=-\lambda\nabla\theta$, the thermodynamic force $\bfp=-\partial\Psi/\partial\bfupsilon$, and the boundary conditions
\be
\begin{cases}
\D\bfn\cdot\bfsigma = \bft^\star \\[3mm]
\D\lambda\bfn\cdot\nabla\theta = \gamma^\star \\[3mm]
\D \bfn\cdot\nabla\bfupsilon = \boldsymbol{0}
\end{cases} \qquad\forall\ \bfx\in\partial\Omega \ .
\ee

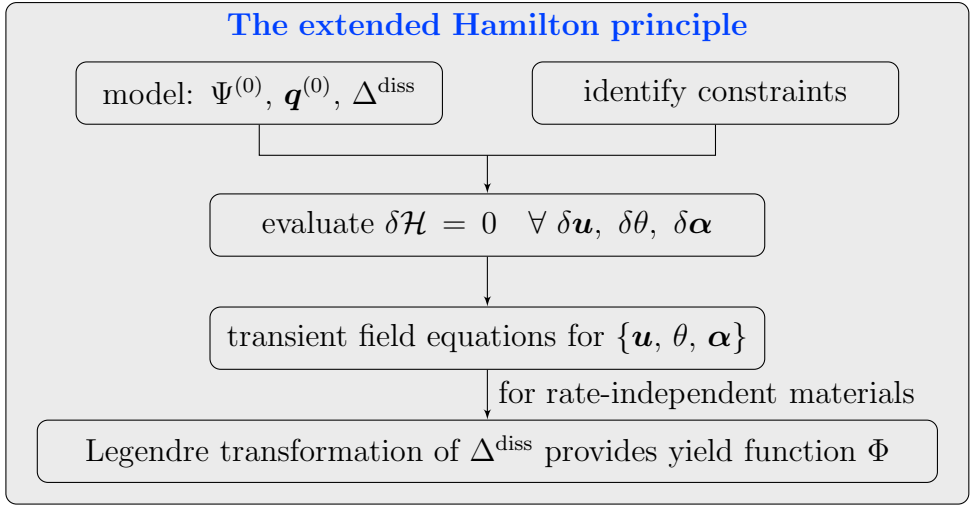
\begin{figure}
\centering
    \tikzstyle{blind} = [circle, 
    text width=0em, text centered, minimum height=0em]
\tikzstyle{line} = [draw, -latex']
\tikzstyle{ini1} = [rectangle, draw, 
    text width=30em, text centered, rounded corners, minimum height=2em,fill=lightgray]
\tikzstyle{ini2} = [rectangle, draw, 
    text width=11em, text centered, rounded corners, minimum height=2em,fill=lightgray]
\tikzstyle{ini3} = [rectangle, draw, 
    text width=17em,   text centered, rounded corners, minimum height=2em,fill=lightgray]
\tikzstyle{ini4} = [rectangle, draw, 
    text width=28em,   text centered, rounded corners, minimum height=2em,fill=lightgray]        
\tikzstyle{ini5} = [rectangle, draw, 
    text width=20em,   text centered, rounded corners, minimum height=2em,fill=lightgray]     
    \begin{tikzpicture}[node distance = 2.5cm, auto]
    \node [ini1] (Hamilton) at (0,0.375) {\textbf{\textcolor{blau}{The extended Hamilton principle}}\vspace{6.0cm}   
	};
    \node[ini2](Hmodel) at (-3,2.5) { 
    model: $\Psi^\refc$, $\heatflux^\refc$, $\Pidissrefc$
    };
    \node[ini2](Hcons) at (3,2.5) { 
   identify constraints
    };    
    \node[ini3](Hevaluate) at (0,0.75) {
    evaluate $\delta\calH = 0 \quad\forall \ \delta\bfu, \ \delta\theta, \ \delta \bfupsilon$
    };
    \node[ini3](Hresult) at (0,-0.75) {
   transient field equations for $\{\bfu, \,\theta, \, \bfupsilon\}$
    };
    \node[ini4](Hyield) at (0,-2.25) {
  Legendre transformation of $\calD^\text{diss}$ provides yield function $\Phi$
    };
	\draw  (Hmodel) -- (-3,1.675) -- (3,1.675) -- (Hcons);    
	\draw[line] (0,1.675) -- (Hevaluate);
	\draw[line] (Hevaluate) -- (Hresult);
	\draw[line] (Hresult) -- (Hyield)  node[midway] (TextNode1) {for rate-independent materials};
    \end{tikzpicture}
\caption{
General use of extended Hamilton principle: the free energy density $\Psi^\refc$, the heat flux vector $\heatflux^\refc$, and the dissipation function $\Pidissrefc$ have to specified along with possible constraints. Evaluation of extended Hamilton principle $\delta\calH=0$ yields the set of transient field equations for all state variables. 
}
\label{fig:FlowH}
\end{figure}

The general use of classical approaches and Hamilton's principle is depicted in Figs. \ref{fig:FlowC} and \ref{fig:FlowH}, respectively.

\FloatBarrier

\section{Derivation of the stationarity conditions}
\label{sec:Derivations}
We show in this section that the field equations (\ref{eq:FieldEquationU},\ref{eq:FieldEquationT},\ref{eq:FieldEquationA}) follow after detailed evaluation of the stationarity conditions in \eqref{eq:GateauxAll}. To this end, we consider the functional dependency $\bar\Psi:=\bar\Psi(\bfC(\nablaX\bfu),\theta,\bfupsilon,\nablaX\bfupsilon)$ for all derivations which follow. 

\subsection{Stationarity with respect to the displacements}
The stationarity condition with respect to $\bfu$ in \eqref{eq:GateauxAll}$_1$ is given by
\bea
\delta_{\bfu}\calH &=& \int_\tau \Big( \intO \rho^\refc \, \dot{\bfu}\cdot\delta\dot{\bfu} \dd V -  \notag \\
&& \intO \rho^\refc\,\pf{\bar{\Psi}}{\bfC}:\delta\bfC \dd V + \intPO \bft^{\star\,\refc} \cdot\delta\bfu \dd A + \intO \bfb^{\star\,\refc} \cdot\delta\bfu \dd V \Big) \dd t = 0\ .
\eea
The first and the second integral can be evaluated by integration by parts, yielding
\be
\int_\tau \intO \rho^\refc \, \dot{\bfu}\cdot\delta\dot{\bfu} \dd V = \intO \rho^\refc \, \dot{\bfu}\cdot\delta\bfu\Big|_{t_0}^{t_1} \dd V - \int_\tau \intO \rho^\refc\, \ddot{\bfu} \cdot\delta\bfu \dd V 
\ee
and
\bea
\D \intO \rho^\refc\, \pf{\bar{\Psi}}{\bfC}:\delta\bfC \dd V & =& \intO \bfS:\frac{1}{2}\delta\bfC \dd V  = \intO \bfP : \delta\bfF \dd V = \intO \bfP^\text{T} : \nablaX\delta\bfu \dd V \notag\\ 
&  = &\D  \intPO \bfn^\refc\cdot \rev{19}{\bfP^\text{T}} \cdot\delta\bfu \dd A - \intO \nablaX \cdot\rev{19}{\bfP^\text{T}} \cdot\delta\bfu \dd V \ .
\eea
As done for the principle of stationary action, we assume that the displacement at the arbitrary time points $t_0$ and $t_1$ that define the interval over which Hamilton's functional is evaluated are fixated and thus vanishing variations are present, i.e., $\delta\bfu|_{t_0}^{t_1}=\boldsymbol{0}$. Thus, we find
\be
\delta_{\bfu}\calH = \intO \Big( - \rho^\refc\, \ddot{\bfu} + \nablaX\cdot\rev{19}{\bfP^\text{T}} + \bfb^{\star\,\refc} \Big) \cdot\delta\bfu \dd V - \intPO \Big( \bfn^\refc\cdot\rev{19}{\bfP^\text{T}} - \bft^{\star\,\refc} \Big) \cdot\delta\bfu \dd A \ .
\ee
Surface and volume are independent for which the conditions have to hold for surface and volume integrals separately. Consequently, we find for $\delta_{\bfu}\calH=0$
\be
\begin{cases} \D\intO \left(\nablaX\cdot\rev{19}{\bfP^\text{T}} + \bfb^{\star\,\refc} \right) \cdot\delta\bfu \dd V = \intO \rho^\refc \, \ddot{\bfu}\cdot\delta\bfu \dd V\\[3mm]
\D \intPO \left(\bfn^\refc\cdot\rev{19}{\bfP^\text{T}} - \bft^{\star\,\refc}\right)\cdot\delta\bfu \dd A = 0  \end{cases}
\ee
which is exactly \eqref{eq:FieldEquationU}. 

\subsection{Stationarity with respect to temperature} 
The stationarity condition $\delta_\theta \calH$ in \eqref{eq:GateauxAll}$_2$ is computed as
\bea
\D\delta_\theta \calH &=& \D - \intO \rho^\refc\,\pf{\bar{\Psi}}{\theta}\delta\theta \dd V -\intO \int \rho^\refc\,\bar{s} \,\delta\dot{\theta} \dd t \dd V \notag \\
\label{eq:VarHTheta}
&& \D  + \int_\Omega\int \frac{1}{\theta^2} \heatflux^\refc\cdot\nablaX\theta \ \delta\theta \dd t \dd V - \int_\Omega \int \frac{1}{\theta} \heatflux^\refc\cdot \nablaX\delta\theta \dd t \dd V  - \intO \int \ddDsrefc\delta\theta \dd t \dd V = 0
\eea
where we evaluated the variation for fixated heat flux vector and entropy production implying $\delta_\theta\dot{\bfq}^\refc=\boldsymbol{0}$ and $\partial\dissrefc/\partial\theta=0$.

\rev{A}{\textit{Remark:} the purpose of entropy is the consideration of path-dependence by ``constraining'' the physical processes to evolve only in an irreversible manner. This is expressed in terms of the entropy balance and the dissipation inequality, cf.~\eqref{eq:ClausiusDuhem}. The current value of entropy can be either computed by evaluation of the constitutive equation~\eqref{eq:ConstLawEntropy} once the state variables $\bfu$, $\theta$ and $\bfalpha$ are known or from the \emph{process} the material and its microstructure have gone through~\eqref{eq:BalEntropy}. However, since we make use of the state variables $\bfu$, $\theta$ and $\bfalpha$, the entropy is always a \emph{consequence} of the dissipative processes. To be more precise, entropy does not adjust itself such that some energetic quantity would be stationary (in contrast to the strains which minimize the free energy in a simple elastic and iso-thermal setting). Consequently, although possessing a functional dependency on the temperature, we freeze the entropy for the computation of the variation of $\calH$ with respect to temperature, i.e., $\delta_\theta \bar{s} = 0$.}

\rev{A}{\textit{Remark:} the entropy production $\ddDsrefc = \bfpdissrefc\cdot\dot{\bfupsilon}$, cf.~\eqref{eq:DefDiss}, might have a functional dependency on the temperature through $\bfpdissrefc=\bfpdissrefc(\theta)$. However, this functional dependency is not taken into account for the computation of the stationarity: the non-conservative forces $\bfpdissrefc$ do not adjust themselves to turn $\calH$ stationary with respect to temperature. In contrast, they depend on the processes that evolve in the material. Therefore, the non-conservative forces are determined by the material properties, e.g. rate-dependent or rate-independent, and have to be modeled in analogy to the mechanical forces which neither adjust themselves to let $\calH$ become stationary (with respect to $\bfu$). }

It is convenient to eliminate all derivatives of $\delta\theta$, i.e., $\nablaX\delta\theta$ and $\delta\dot{\theta}$, by performing integration by parts in space and time, respectively. We thus compute
\be
-\intO \int \rho^\refc\,\bar{s}\, \delta\dot{\theta} \dd t \dd V = - \intO \rho^\refc\, \bar{s}\, \delta\theta \dd V  + \intO \int \rho^\refc\, \dot{\bar{s}}\, \delta\theta \dd t \dd V \ ,
\ee
where we used mass conservation as $\dot{\rho}^\refc=0$, and
\bea
- \intO \frac{1}{\theta} \heatflux^\refc\cdot\nablaX \delta\theta \dd V &=& - \intPO \frac{1}{\theta}\bfn^\refc\cdot\heatflux^\refc \, \delta\theta \dd A + \intO \nablaX\cdot\Big(\frac{\heatflux^\refc}{\theta}\Big) \, \delta\theta \dd V \notag \\
&=& - \intPO \frac{1}{\theta} \bfn^\refc\cdot\heatflux^\refc \, \delta\theta \dd A \notag \\
&&  - \intO \frac{1}{\theta^2} \heatflux^\refc \cdot \nablaX\theta \, \delta \theta \dd V + \intO \frac{1}{\theta} \nablaX\cdot\heatflux^\refc \, \delta\theta \dd V \ .
\eea
Inserting these results into \eqref{eq:VarHTheta} yields
\bea
\delta_\theta\calH &=& - \intO \rho^\refc\pf{\bar{\Psi}}{\theta}\delta\theta \dd V - \intO \rho^\refc\, \bar{s} \,\delta\theta \dd V  + \intO \int \rho^\refc\, \dot{\bar{s}} \,\delta\theta \dd t \dd V \notag\\
&&  - \intPO\int \frac{1}{\theta} \bfn^\refc\cdot\heatflux^\refc \, \delta\theta \dd A \dd t + \intO \int \frac{1}{\theta} \nablaX\cdot\heatflux^\refc \, \delta\theta \dd t \dd V   - \intO \int \ddDsrefc\delta\theta \dd t \dd V \ .
\eea
Making use of the constitutive law for entropy \eqref{eq:ConstLawEntropy}, i.e. $\bar{s}=-\partial\bar{\Psi}/\partial\theta$, the stationarity condition simplifies to
\be
\label{eq:deltaHT}
\delta_\theta\calH =  - \intPO\int \frac{1}{\theta} \bfn^\refc \cdot \heatflux^\refc \, \delta\theta \dd A \dd t  + \intO \int \Big( \rho^\refc\, \dot{\bar{s}}   + \frac{1}{\theta} \nablaX\cdot\heatflux^\refc-   \ddDsrefc \Big) \delta\theta \dd t \dd V \ .
\ee
Let us introduce $\Psi^\refc=\rho^\refc\,\bar{\Psi}$ and $s^\refc=\rho^\refc\,\bar{s}$. Then, the rate of entropy is given by
\be
\dot{s}^\refc = - \frac{\mathrm{d}}{\mathrm{d} t} \pf{\Psi^\refc}{\theta} = - \pf{}{\theta} \left[ \pf{\Psi^\refc}{\bfC}:\dot{\bfC} +\pf{\Psi^\refc}{\theta}\dot{\theta} + \pf{\Psi^\refc}{\bfupsilon}\cdot\dot{\bfupsilon} \right] 
\ee
when the constitutive equation $s^\refc =-\partial \Psi^\refc/\partial \theta$ in \eqref{eq:ConstLawEntropy} is used. Considering \rev{28}{the definition of the 2${}^\text{nd}$ Piola-Kirchhoff stress tensor $\bfS$ and the heat capacity $\kappa:=-\theta \partial^2\Psi^\refc/\partial\theta^2$, cf. \cite{altenbach2012kontinuumsmechanik}, 
the rate of entropy can be approximated by
\be
\theta \, \dot{s}^\refc =  - 2 \theta \, \pf{\bfS}{\theta} :\dot{\bfC} + \kappa \, \dot{\theta} - \theta \, \pf{\bfp}{\theta}\cdot\dot{\bfupsilon} \ . 
\ee }
Furthermore, we recall the definition for the entropy production according to \eqref{eq:DefDiss} $\ddDsrefc=\bfpdissrefc\cdot\dot{\bfupsilon}$ which, after consideration of \eqref{eq:Biot1} \re{and \eqref{eq:StatHSimpleMaterial}}, equals $\ddDsrefc=\bfpdissTrefc\cdot\dot{\bfupsilon}/\theta \re{=\bfp\cdot\dot{\bfalpha}/\theta}$. Then, \eqref{eq:deltaHT} transforms to
\be
\begin{cases}
\D\intO \int \frac{1}{\theta} \left( \kappa\dot{\theta} + \nablaX\cdot\heatflux^\refc \re{- 2 \theta \, \pf{\bfS}{\theta} :\dot{\bfC} - \left[\bfp + \theta\,\pf{\bfp}{\theta} \right]}\cdot\dot{\bfupsilon}\right) \delta\theta \dd t \dd V &=0 \quad\forall \bfx \in \Omega \\
\D\intPO \int \frac{1}{\theta} \bfn^\refc \cdot\heatflux^\refc \, \delta\theta \dd t \dd A & = 0 \quad \forall \bfx \in \partial\Omega
\end{cases}
\ee
from which \eqref{eq:FieldEquationT} immediately follows. 

An important special case are isothermal processes for which $\nablaX\theta=\boldsymbol{0}$ and $\delta\theta\not=\delta\theta(t)$ hold. Consequently, we can write for \eqref{eq:VarHTheta}
\be
-\intO \rho^\refc\, \pf{\bar{\Psi}}{\theta} \delta\theta \dd V - \intO \int \ddDsrefc \dd t  \ \delta\theta \dd V = 0
\ee
and thus
\be
\label{eq:sisotherm}
\rho^\refc \bar{s} = s^\refc = \int \ddDsrefc \dd t \ .
\ee

\subsection{Stationarity with respect to internal variables}
We consider non-local effects of the microstructure and, hence, of the internal variable. Consequently, the free energy is assumed to be modeled as gradient-enhanced quantity with respect to $\bfupsilon$ such that the related stationarity condition in \eqref{eq:GateauxAll}$_3$ takes the form
\be
\label{eq:StatHv0}
\delta_{\bfupsilon}\calH = - \intO \rho^\refc\,\pf{\bar{\Psi}}{\bfupsilon} \cdot\delta\bfupsilon \dd V - \intO \rho^\refc\,\pf{\bar{\Psi}}{\nablaX\bfupsilon}:\nablaX\delta\bfupsilon \dd V - \intO \int \delta_{\bfupsilon}(\theta \ddDsrefc) \dd t \dd V = 0\ .
\ee
Integration by parts of the second term results in
\be
 \intO \rho^\refc\,\pf{\bar{\Psi}}{\nablaX\bfupsilon}:\nablaX\delta\bfupsilon \dd V = \intPO \rho^\refc\,\bfn^\refc\cdot \pf{\bar{\Psi}}{\nablaX\bfupsilon} \cdot \delta\bfupsilon \dd A - \intO \rho^\refc\, \nablaX \cdot \pf{\bar{\Psi}}{\nablaX\bfupsilon} \cdot \delta\bfupsilon \dd V
\ee
We recognize that the third term in \eqref{eq:StatHv0} is the only one that includes the antiderivative with respect to time. It is thus convenient to introduce
\be
\int \theta \ddDsrefc \dd t =: D \ .
\ee
The quantity $D$ possesses the physical unit of a volume-specific energy and thus, it is referred to as non-conservative dissipated energy. Modeling $D$ independently to $\theta\ddDsrefc$ violates the fundamental theorem of calculus as intended: hereby, a physically motivated path-dependence is ensured. Consequently, the stationarity condition with respect to microstructure is given by
\be
\delta_{\bfupsilon}\calH = - \intO \Big( \rho^\refc\,\pf{\bar{\Psi}}{\bfupsilon} - \rho^\refc\,\nablaX\cdot \pf{\bar{\Psi}}{\nablaX\bfupsilon} + \pf{D}{\bfupsilon} \Big)\cdot\delta\bfupsilon \dd V -  \intPO \rho^\refc\,\bfn^\refc\cdot \pf{\bar{\Psi}}{\nablaX\bfupsilon} \cdot \delta\bfupsilon \dd A  \ .
\ee
Finally, the stationarity with respect to microstructure remains to be analyzed. To this end, the dissipated energy $D$ needs to be specified. We assume a form similar to the mechanical work and postulate
\be
\label{eq:DefDissiaptedEnergy}
D := \bfpdissTrefc \cdot \bfupsilon \ .
\ee
Thus, dissipated energy is generated while the non-conservative force $\bfpdissTrefc$ acts ``along the microstructure'' $\bfupsilon$. From the perspective of application this implies that the non-conservative force has to be modeled rather than the dissipated energy itself. However, $D$ can be computed once the internal variable is known from solving \eqref{eq:FieldEquationA}. The choice of $\bfpdissTrefc$~has to be in accordance to the microstructure evolution, i.e., it has to cover rate-independent and rate-dependent aspects. Usually, the non-conservative force $\bfpdissTrefc$ is assumed to be derivable from a dissipation function $\calD^\text{diss}$ by
\be
\label{eq:DissipationFunction}
\bfpdissTrefc =: \pf{\calD^{\text{diss}\,\refc}}{\dot{\bfupsilon}}
\ee
which simplifies the modeling of microstructure evolution: in contrast to $\bfpdissTrefc$, the dissipation function $\Pidissrefc$ is a scalar-valued quantity. Consequently, the dissipated energy $D$ in \eqref{eq:DefDissiaptedEnergy} transforms to
\be
\label{eq:DefDissipatedEnergy2}
D= \pf{\Pidissrefc}{\dot{\bfupsilon}}\cdot\bfupsilon \ .
\ee
With the definition for the dissipated energy $D$ in \eqref{eq:DefDissipatedEnergy2} and assuming, in analogy to $\partial\ddDsrefc/\partial\theta=0$, that $\partial\bfpdissTrefc/\partial\bfupsilon=\boldsymbol{0}$, the stationarity condition with respect to microstructure transforms to
\be
\label{eq:DerivationA}
\begin{cases}
\D\intO \Big( \pf{\Psi^\refc}{\bfupsilon} - \nablaX\cdot \pf{\Psi^\refc}{\nablaX\bfupsilon} + \pf{\Pidissrefc}{\dot{\bfupsilon}} \Big)\cdot\delta\bfupsilon \dd V = 0 \\[3mm]
\D \intPO \bfn^\refc\cdot \pf{\Psi^\refc}{\nablaX\bfupsilon} \cdot \delta\bfupsilon \dd A  = 0 
\end{cases}  
\ee
when \eqref{eq:DissipationFunction} has been inserted. Then, \eqref{eq:DerivationA} equals \eqref{eq:FieldEquationA}.

\section{Examples of the extended Hamilton principle in material modeling}
In this section, we show that application of the extended Hamilton principle yields to physically reasonable governing equation. It will be obvious that no ``novel'' governing equations are obtained; in contrast, we receive the same material models as already established in literature. However, they result here from the stationarity conditions of the extended Hamilton principle.
\re{\subsection{Thermo-elastic materials for small deformations}}
\rev{28}{
As first example, we apply the formulas to thermo-elastic materials for small deformations. Here, no internal variable is needed such that $\bfalpha\equiv\boldsymbol{0}$. The free energy density is given by
\be
\Psi = \frac{1}{2} (\bfeps - \alpha^\text{th} \theta \bfI):\dsC:(\bfeps - \alpha^\text{th} \theta \bfI)
\ee
with the thermal expansion coefficient $\alpha^\text{th}$ and the elasticity tensor $\dsC$. Evaluation of the stationarity conditions of the extended Hamilton functional then yields
\be
\begin{cases}
\nabla\cdot\bfsigma + \bfb^\star = \rho \ddot{\bfu} \\
\kappa \, \dot{\theta} = - \alpha^\text{th} \, \theta \; \mathrm{tr}[\dsC:\dot{\bfeps} ] + \lambda \nabla\cdot\nabla\theta
\end{cases} \forall \ \bfx\in\Omega
\ee
with the boundary conditions
\be
\begin{cases}
\bfn\cdot\bfsigma = \bft^\star \\
\bfn\cdot\nabla\theta = 0
\end{cases} \forall \ \bfx\in\partial\Omega
\ee
and $\mathrm{tr}[\dsC:\dot{\bfeps}]=\bfI:\dsC:\dot{\bfeps}$.
}

\subsection{Visco-elastic materials for small deformations}
In case of visco-elastic materials with linearized kinematics, the internal variable is the viscous part of the total strain which is denoted by $\bfepsV$. \re{Here and in the following, we neglect thermo-elastic coupling since the heat production due to dissipative processes is usually much higher.} The free energy density is given by 
\be
\Psi^\text{mech}=\frac{1}{2}(\bfeps-\bfepsV):\dsC:(\bfeps-\bfepsV) \ .
\ee
For the rate-dependent microstructure evolution, we use
\be
\calD^{\text{diss}\,\text{v}} = \frac{\eta}{2} ||\dbfepsV||^2 \ .
\ee
The constraint of volume preservation is given in terms of the rates of the viscous strains by
\be
\label{eq:ConsVisco}
\mathrm{tr}\dbfepsV = 0 
\ee
which is accounted for by expanding the extended Hamilton functional as
\be
\calH^\text{v} = \calH + \int_\tau \intPO \int \tilde{\gamma}^\star \theta \dd t \dd A \dd t + \int_\tau \int_\Omega g \, \mathrm{tr}\dbfepsV  \dd V \dd t 
\ee
with a Lagrange parameter $g$. Note that the second term accounts for prescribed heat flux at the surface.

Application of the extended Hamilton principle, following Fourier's law and inserting the constraint \eqref{eq:ConsVisco} yields the following set of equations:
\be
\begin{cases}
\D\nabla\cdot\bfsigma + \bfb^\star = \rho \, \ddot{\bfu} \\[1mm]
\D \kappa \, \dot{\theta} = \mathrm{dev}\bfsigma:\dbfepsV \re{+} \lambda \nabla\cdot\nabla\theta \\[1mm]
\D \dbfepsV = \frac{1}{\eta} \mathrm{dev}\bfsigma
\end{cases} \quad \forall \ \bfx\in\Omega
\ee
with the boundary conditions
\be
\begin{cases}
\bfn\cdot\bfsigma = \bft^\star \\[1mm]
\lambda \bfn\cdot\nabla\theta = \gamma^\star
\end{cases} \forall \ \bfx\in\partial\Omega \ .
\ee
It is worth mentioning that the stress deviator
\be
\mathrm{dev}\bfsigma = \bfsigma -  \bfI \, \mathrm{tr}\bfsigma
\ee
is a result of the stationarity condition due to the constraint and was not defined in advance.

\subsection{Finite elasto-plasticity}
In case of finite elasto-plasticity, we follow the standard kinematics by introducing an intermediate configuration and the multiplicative split
\be
\bfF = \bfF^\el\cdot\bfF^\pl
\ee
into an elastic and plastic part of the deformation gradient $\bfF^\el$ and $\bfF^\pl$, respectively, see \cite{kroner1959allgemeine,lee1969elastic}. The internal variable is in this case the amount of plastic deformation which we measure by the so-called plastic variable
\be
\bfFp := \bfF^{\pl\,-1}
\ee
see e.g. \cite{carstensen2002non}. 
The free energy density is given by $\bar{\Psi}=\bar{\Psi}(\bfF^\el)=\bar{\Psi}(\bfF,\bfFp)$. Consequently, the 1${}^\text{st}$ Piola-Kirchhoff stress tensor reads
\be
\bfP = \rho^\refc\pf{\bar{\Psi}}{\bfF} =  2 \rho^\refc\bfF\cdot\pf{\bar{\Psi}}{\bfC} =  \rho^\refc\pf{\bar{\Psi}}{\bfF^\el} \cdot \bfFp{}^T \ .
\ee
The thermodynamic driving force is
\be
\bfp = -\rho^\refc\pf{\bar{\Psi}}{\bfFp} = -\rho^\refc \pf{\bar{\Psi}}{\bfF^\el}:\pf{\bfF^\el}{\bfFp} =  -\rho^\refc \bfF^T\cdot\pf{\bar{\Psi}}{\bfF^\el} \ .
\ee
The dissipation function is modeled by
\be
\calD^{\text{diss}\,\pl} = r^\mathrm{p} ||\bfF^\pl \cdot \dbfFp||
\ee
The motivation for including the plastic deformation gradient $\bfF^\pl$ will be discussed later. 
The rate of the plastic variable is given by
\be
\dbfFp = - \bfFp\cdot\dot{\bfF}{}^\pl\cdot\bfFp \ .
\ee
Then, the extended Hamilton functional is used as
\be
\calH^\text{p} = \calH + \int_\tau \intPO \int \tilde{\gamma}^\star \theta \dd t \dd A \dd t \ . 
\ee
The stationarity conditions of Hamilton's functional demand
\be
\begin{cases}
\D\nablaX\cdot\rev{19}{\bfP^\text{T}} + \bfb^{\star\,\refc} = \rho^\refc \, \ddot{\bfu} \\[1mm]
\D \kappa \, \dot{\theta} = \bfM:\bfL^\pl \re{+} \lambda \nablaX\cdot\nablaX\theta \\[1mm]
\D \bfL^\pl \ni \frac{||\bfF^\pl\cdot\dbfFp||}{r^\mathrm{p}}\bfM =: \re{k}\, \bfM
\end{cases} \forall \ \bfx\in\Omega
\ee
where $\bfL^\pl:=\dot{\bfF}^\pl\cdot\bfFp$ is the plastic velocity gradient and $\bfM:=\rho^\refc\,\bfF^\el{}^T\cdot\partial\bar{\Psi}/\partial\bfF^\el$ is the Mandel stress tensor. Again, we used Fourier's law for modeling $\heatflux^\refc$. The boundary conditions follow to be
\be
\begin{cases}
\bfP\cdot\bfn^\refc = \bft^{\star\,\refc} \\[1mm]
\lambda \rev{32}{\bfn^\refc}\cdot\nablaX\theta = \gamma^\star
\end{cases} \forall \ \bfx\in\partial\Omega \ .
\ee
The quantity $\re{k}$ is the consistency parameter. Performing a Legendre transformation of the dissipation function results in the yield function and associated Karush-Kuhn-Tucker conditions
\be
\re{k} \ge 0 \ , \qquad \Phi^\text{p} := ||\bfM|| - r^\mathrm{p} \le 0 \ , \qquad \re{k}\,\Phi^\text{p} = 0 \ .
\ee
The parameter $r$ is thus identified as the norm of the yield stress. Comparable plasticity models can be found, e.g. in \cite{hackl1997generalized,carstensen2002non}. Inspection of the evolution equation reveals that with the used choice for $\calD^{\text{diss}\,\pl}$, i.e. consideration of $\bfF^\pl$, all dependencies on the plastic deformation vanish and thus that the model is invariant under all plastic deformations.

\subsection{Gradient-enhanced damage modeling for small deformations}
The internal variable is the scalar-valued damage variable $d$ which enters the damage function $f(d)=e^{-d}$. The gradient-enhanced free energy density is given by
\be
\Psi^\text{mech} = f \Psi^\text{mech}_{\rev{33}{0}} + \frac{1}{2} \beta ||\nabla f||^2 \qquad\text{with}\qquad \Psi^\text{mech}_0 = \frac{1}{2}\bfeps:\dsC:\bfeps 
\ee
\rev{27}{and the regularization parameter $\beta:= \ell^2 \, E >0$ which has the physical unit Newton. Here, $E$ denotes the Young's modulus and $\ell$ is the internal length.} The dissipation function is chosen to be rate-independent, thus
\be
\calD^{\text{diss}\,\mathrm{d}} = r^\mathrm{d} |\dot{d}| \ .
\ee
The extended Hamilton functional is
\be
\calH^\text{d} = \calH + \int_\tau \intPO \int \tilde{\gamma}^\star \theta \dd t \dd V \dd t 
\ee
such that the stationarity conditions result in 
\be
\begin{cases}
\D\nabla\cdot\bfsigma + \bfb^\star = \rho \, \ddot{\bfu} \\[1mm]
\D \kappa \, \dot{\theta} = f\Psi_0^{\rev{34}{\text{mech}}} \, \dot{d} \re{+} \lambda \nabla\cdot\nabla\theta \\[1mm]
\D r^\mathrm{d} \frac{\dot{d}}{|\dot{d}|} \ni f\Psi_0^{\rev{34}{\text{mech}}} - \beta f \nabla\cdot\nabla f
\end{cases} \forall \ \bfx\in\Omega
\ee
with  the boundary conditions
\be
\begin{cases}
\bfn\cdot\bfsigma = \bft^\star \\[1mm]
\lambda \bfn\cdot\nabla\theta = \gamma^\star \\[1mm]
\beta \bfn\cdot\nabla f = 0
\end{cases} \forall \ \bfx\in\partial\Omega \ .
\ee
Analyzing the Legendre transformation of the dissipation function $\calD^{\text{diss}\,\mathrm{d}}$ results in the Karush-Kuhn-Tucker conditions
\be
\dot{d} \ge 0 \ , \qquad \Phi^\text{d} := f\Psi_0 - \beta f \nabla\cdot\nabla f - r^\mathrm{d} \le 0 \ , \qquad \dot{d}\,\Phi^\text{d} = 0 
\ee
with the energetic threshold value for damage initiation and evolution $r^\mathrm{d}$. An isothermal version of the model was presented in \cite{junker2019fast}.

\section{Conclusions}
In this paper, using the example of thermo-mechanically coupled processes, we showed that Hamilton's principle can be extended to provide a unifying theory for coupled problems and dissipative microstructure evolution. To this end, we presented a brief overview on the history of material modeling and recalled the fundamentals of thermodynamics. We continued with the principle of stationary action as governing axiom for an energy-based description of the behavior of rigid particles. Starting from the principle of stationary action for non-conservative rigid particles, we presented a generalization which is valid for deformable solids. Combination of the balance of energy and the action functional for continuous bodies resulted in an extended Hamilton functional. By postulating stationarity of Hamilton's functional with respect to all state variables, we obtained Euler-Lagrange equations for the displacements, temperature, and internal variables. The Euler-Lagrange equations covered all governing equations for solids, including the balance of linear momentum, Cauchy's theorem, the principle of virtual work, heat equation, surface heat flux, Biot's equation, evolution equations with and without non-local contributions, and the principle of the minimum of the dissipation potential. Consequently, we showed that Hamilton's principle unifies the thermo-mechanical field equations from a variational perspective. For demonstration purposes, we presented the thermo-mechanically coupled governing equations resulting from the extended Hamilton principle for \re{four} exemplary materials: \re{linearized thermo-elasticity,} linearized visco-elasticity, finite plasticity, and rate-independent, gradient-enhanced brittle damage modeling. For other couplings such as e.g., electro-mechanical or chemo-mechanical problems, the analogous analysis remains as open problem for future research.


\addcontentsline{toc}{chapter}{Bibliography}
\bibliographystyle{plain}
\bibliography{bib_short}

\end{document}